\documentclass[12pt, notitlepage]{article}

\usepackage{sectsty}
\allsectionsfont{\sffamily}

\usepackage{a4}
\usepackage{amssymb, graphicx,psfrag,epsf}
\usepackage{enumerate}
\usepackage{subfigure}
\usepackage{amsmath}
\usepackage{amsthm}
\usepackage{bbm}
\usepackage{undertilde}
\usepackage{verbatim}
\usepackage{bbm}
\usepackage{ dsfont }
\usepackage{geometry}
\usepackage{algorithm} 
\usepackage[noend]{algpseudocode}
\usepackage{multirow}
\usepackage{pdflscape}
\usepackage{setspace}
\usepackage{hyperref}
\hypersetup{
	colorlinks=true,
	linkcolor=TUMblue,
	citecolor=TUMblue,
	filecolor=TUMblue,
	urlcolor=TUMblue
}
\usepackage{array}
\usepackage{xcolor}
\usepackage{bm}
\usepackage{cleveref}

\definecolor{TUMblue}{RGB}{0, 101, 189}
\definecolor{TUMlightblue}{RGB}{100,160,200}
\definecolor{TUMgreen}{RGB}{162,173,0}
\definecolor{TUMorange}{RGB}{227,114,034}
\definecolor{TUMivory}{RGB}{218,215,203}

\usepackage{aliascnt}
\usepackage{makecell}

\newcommand{\mynewtheorem}[2]{
	\newaliascnt{#1}{dummy}
	\newtheorem{#1}[#1]{#2}
	\aliascntresetthe{#1}
	\expandafter\def\csname #1autorefname\endcsname{#2}
}

\theoremstyle{definition}
\mynewtheorem{thm}{Theorem}
\mynewtheorem{defi}{Definition}
\mynewtheorem{lem}{Lemma}
\mynewtheorem{cor}{Corollary}
\mynewtheorem{prop}{Proposition}
\mynewtheorem{exa}{Example}
\mynewtheorem{alg}{Algorithm}
\mynewtheorem{rem}{Remark}
\mynewtheorem{bsp}{Example}

\def\equationautorefname~#1\null{Equation~(#1)\null}

\newcolumntype{M}{>{\centering\arraybackslash}m{}}

\usepackage{url}
\usepackage{natbib}

\newcommand{\N}{\mathbb{N}}

\newcommand{\Ub}{\mathbf{U}}
\newcommand{\ub}{\mathbf{u}}

\newcommand{\0}{\mathbf{0}}

\newcommand{\be}{\begin{equation}}
\newcommand{\ee}{\end{equation}}

\DeclareMathOperator{\rank}{rank}

\begin{document}

	\title{\bf Growing simplified vine copula trees:\\ improving Di\ss mann's algorithm}
	\date{\small \today}
	\author{Daniel Kraus\footnote{Corresponding author, Zentrum Mathematik, Technische Universit\"at M\"unchen, Boltzmanstra\ss e 3, 85748 Garching (email: \href{mailto:daniel.kraus@tum.de}{daniel.kraus@tum.de})}\hspace{.2cm} and Claudia Czado\footnote{Zentrum Mathematik, Technische Universit\"at M\"unchen, Boltzmanstra\ss e 3, 85748 Garching}}
	\maketitle

\vspace{-3mm}
  \begin{abstract}
Vine copulas are pair-copula constructions enabling multivariate dependence modeling in terms of bivariate building blocks. One of the main tasks of fitting a vine copula is the selection of a suitable tree structure. For this the prevalent method is a heuristic called Di\ss mann's algorithm. It sequentially constructs the vine's trees by maximizing dependence at each tree level, where dependence is measured in terms of absolute Kendall's $\tau$. However, the algorithm disregards any implications of the tree structure on the simplifying assumption that is usually made for vine copulas to keep inference tractable. We develop two new algorithms that select tree structures focused on producing simplified vine copulas for which the simplifying assumption is violated as little as possible. For this we make use of a recently developed statistical test of the simplifying assumption. In a simulation study we show that our proposed methods outperform the benchmark given by Di\ss mann's algorithm by a great margin. Several real data applications emphasize their practical relevance. 

  \end{abstract}

\section{Introduction}\label{sec:introduction}

The list of references in which vine copulas are used to describe the dependence structure of multivariate random variables is long and does not cease to grow. After their inauguration in \cite{aas2009pair} these pair-copula constructions have been applied to statistical problems in biology \citep{barthel2016vine,schellhase2016estimating}, sociology \citep{cooke2015vine}, hydrology \citep{erhardt2015standardized,hobaek2015well,killiches2016block,pereira2016spatial} and finance \citep{maya2015latin,almeidamodeling,kraus2017d}, to only name a few recent ones. An extensive review of vine copulas and their applications in finance is given in a survey by \cite{aas2016pair}. In order to illustrate the idea of vine copulas we present an exemplary vine copula decomposition of a three-dimensional vine copula:
\begin{equation}\label{eq:vine3d}
c(u_1,u_2,u_3)=c_{12}(u_1,u_2)\,c_{23}(u_2,u_3)c_{13;2}\left(C_{1|2}(u_1|u_2),C_{3|2}(u_3|u_2);u_2\right).
\end{equation}
The multivariate copula density is expressed in terms of bivariate building blocks called pair-copulas. Two of them are unconditional ($c_{12}$ and $c_{23}$) and one is conditioned ($c_{13;2}$ on $u_2$). In general, there are various ways of specifying a vine copula decomposition. E.g.\ in the three-dimensional case, there are three possibilities depending on which variable is chosen to be the conditioning one. When we speak of the structure of the vine, we mean the specification which pairs of variables are modeled conditioned on which other variables.\\
It is now a stylized fact that the flexibility of vines is able to overcome the problem of a limited model choice in the case when only elliptical or Archimedean copulas are used. Being able to assign a different bivariate copula to any pair of variables, the new issue arises of having too many possible models to choose from. Not only the selection of pair-copula families and corresponding parameters, but especially the vast number of possible model structures make the search of an ``optimal'' vine for a given data set a challenging task. Until now, almost unanimously, the method of choice for structure selection is the so-called Di\ss mann algorithm \citep{dissmann2013selecting}. This is also due to its prominent implementation as function \texttt{RVineStructureSelect} in the R-package \texttt{VineCopula} \citep{VineCopula}. Di\ss mann's algorithm is a heuristic that sequentially constructs the vine's structure, trying to capture most of the dependence in the lower trees.\\
When fitting a vine copula we usually make the so-called \textit{simplifying assumption}, i.e.\ that the bivariate copulas of conditioned pairs of variables in the vine copula model do not depend on the value of the variable(s) they are conditioned on. In the three-dimensional example of \autoref{eq:vine3d} this would imply that the conditional copula density $c_{13;2}$ would not depend on $u_2$. The implications and possible relaxation of this assumption have been discussed thoroughly in the recent literature \citep[][and many others]{killiches2016examination,vatter2016generalized,spanhel2015simplified,acar2012beyond}. Even though the simplifying assumption is a model restriction, it is the only way to make inference for vine copulas tractable in higher dimensions.\\
The motivation of this paper is the observation that the model structure of a vine copula has a considerable influence on the validity of the simplifying assumption. To illustrate this, reconsider the three-dimensional vine copula model from \autoref{eq:vine3d}, now in its simplified form:
\[
c(u_1,u_2,u_3)=c_{12}(u_1,u_2)\,c_{23}(u_2,u_3)c_{13;2}\left(C_{1|2}(u_1|u_2),C_{3|2}(u_3|u_2)\right).
\]
This density can also be written as another vine copula using a different tree structure, e.g.\ the one containing the pairs $(1,2)$ and $(1,3)$ in the first and $(2,3;1)$ in the second tree:
\[
c(u_1,u_2,u_3)=c_{12}(u_1,u_2)\,c_{13}(u_1,u_3)c_{23;1}\left(C_{2|1}(u_2|u_1),C_{3|1}(u_3|u_1);u_1\right).
\]
This decomposition does not need to be of simplified form anymore, i.e.\ the copula $c_{23;1}$ might now depend on the conditioning value $u_1$. Hence, given data from the simplified vine copula model we expect a fitted simplified vine copula with the original tree structure to have a significantly better model fit than one with a different tree structure. While Di\ss mann's structure selection algorithm only takes into account the strength of dependencies between pairs of variables, no consideration about the resulting conditional pair-copula is made. In this paper, we propose two new structure selection algorithms that try to amend exactly this flaw. By using a test for constant conditional correlations developed in \cite{kurz2017pacotest}, we incorporate information about the violation of the simplifying assumption when constructing the vine's tree structure. The first algorithm selects the first tree similar to Di\ss mann's algorithm and uses the p-values of these tests as weights in the subsequent trees. The second algorithm we propose fits a C-vine where each root node is selected in a way that minimizes the occurrence of non-simplifiedness in the next tree. As we will see in a simulation study, this improves the model fit in terms of the Akaike information criterion (AIC) compared to Di\ss mann most of the time, especially when the dimension is large. Finally, in revisiting several classic data sets that have already been studied in the vine copula context we demonstrate the practical relevance of our proposed methods.\\
The remainder of the paper is organized as follows. \autoref{sec:theory} provides the theoretical background needed to understand the concepts introduced in this paper. Two new algorithms for a vine's structure selection are motivated and proposed in \autoref{sec:algorithms}. The performance of these algorithms is evaluated in a simulation study in \autoref{sec:simstudy}. Real data applications are given in \autoref{sec:realdata} and \autoref{sec:conclusion} concludes.

\section{Theoretical Background}\label{sec:theory}

\subsection{Vine copulas}
We shortly recall the most important definitions needed for the construction of vine copulas. Details can be found in \cite{bedford2002vines} or \cite{aas2009pair}. Moreover, we assume the concept of copulas to be known (refer to \cite{nelsen2007introduction} and \cite{joe1997multivariate} for a thorough introduction to the topic) and restrict all our analyses to variables and data on the uniform copula scale $[0,1]^d$. 

A $d$-dimensional vine copula is a pair-copula construction consisting of $d(d-1)/2$ unconditional and conditional bivariate copulas, whose structure is organized by a set of linked trees $\mathcal{V}=(T_1,\ldots,T_{d-1})$ satisfying
\begin{enumerate}
	\item[(i)] $T_1=(V_1,E_1)$ is a tree with nodes $V_1=\left\{1,\ldots,d\right\}$ and edges $E_1$. A tree is understood as a graph where any two nodes are connected by a unique	path \citep[refer to][for an introduction to graph theory]{diestel2000graduate}.
	\item[(ii)] For $m=2,\ldots,d-1$, the tree $T_m$ consists of nodes $V_m=E_{m-1}$ and edges $E_m$.
	\item[(iii)] For $m=2,\ldots,d-1$, two nodes of $T_m$ can only be connected by an edge if the corresponding edges of $T_{m-1}$ have a common node.
\end{enumerate}

Each edge $e$ of the vine copula model's $d-1$ trees is associated with a bivariate pair-copula $c_{j_e,k_e;D_e}$, where -- following the notation of \cite{czado2010pair} -- $j_e$ and $k_e$ denote the indices of the conditioned variables $U_{j_e}$ and $U_{k_e}$, and $D_e$ represents the conditioning set corresponding to edge $e$. Thus, $c_{j_e,k_e;D_e}$ is the density of the copula between random variables $U_{j_e|D_e}$ and $U_{k_e|D_e}$, where $U_{i|D}:=C_{i|D} (U_i |U_D)$. The vine density can then be written as
\begin{equation}\label{eq:vine}
c(u_1,\ldots,u_d)=\prod_{m=1}^{d-1}\prod_{e\in E_m}c_{j_ek_e;D_e}\left(C_{j_e|D_e}(u_{j_e}|\ub_{D_e}), C_{k_e|D_e}(u_{k_e}|\ub_{D_e});\ub_{D_e} \right). 
\end{equation}

Condition (iii) for the set of trees $\mathcal{V}$ is also known as the \textit{proximity condition}. When one constructs the trees sequentially, starting with $T_1$, then it induces constraints on the nodes in the next tree which can be connected by an edge. Two types of trees play a special role regarding the proximity condition. For $m=1,\ldots,d-2$, if tree $T_m$ has a star like structure, i.e.\ there is one node $j$ connected to all the other nodes, then the proximity condition imposes no restriction for the construction of tree $T_{m+1}$, since all nodes share the common node $j$ from $T_m$. Thus, all nodes of tree $T_{m+1}$ are allowed to be connected. On the contrary, if tree $T_m$ has a path like structure, i.e.\ each node has at most two neighbors, then all higher trees $T_{m+1},\ldots,T_{d-1}$ are already determined by the proximity condition, since for every node in $T_{m+1}$ there exist at most two nodes that can be connected to it, resulting in a path like structure for tree $T_{m+1}$ as well. A vine structure where all trees consist of paths is called a \textit{D-vine}, while a vine with only star like trees is known as a \textit{C-vine}.

\subsection{Di\ss mann's algorithm for selecting simplified vine copulas}\label{sec:dissmann}
Since the sequential structure selection algorithm proposed in \cite{dissmann2013selecting} will serve as a benchmark vine model selection method, we shortly outline how it works. Assume we are given an i.i.d.\ sample of size $n$ of a $d$-dimensional copula, denoted by $(u^{(i)}_j)_{j=1,\ldots,d}^{i=1,\ldots,n}$. Since a vine copula model consists of three parts (the vine structure, the parametric families of each pair-copula and their corresponding copula parameters), each of these components have to be estimated.\\
Assume at first that the tree structure is given. Then for each of the $d(d-1)/2$ pair-copulas a family and copula parameter have to be selected. Starting with the first tree, for each unconditional pair the maximum likelihood estimate of the copula parameter is determined for each family studied (see \autoref{app:families} for a list of families currently implemented in the \texttt{VineCopula} package). Then the family with the lowest Akaike information criterion (AIC) is chosen \citep{akaike1973information}. In the higher trees, for each edge corresponding to a conditional pair-copula $c_{j_e,k_e;D_e}$, the estimated pseudo observations $u^{(i)}_{j_e|D_e}:=\hat C_{j_e|D_e} (u^{(i)}_{j_e}|\ub^{(i)}_{D_e})$ and $u^{(i)}_{k_e|D_e}:=\hat C_{k_e|D_e} (u^{(i)}_{k_e}|\ub^{(i)}_{D_e})$, $i=1,\ldots,n$, are calculated and used to find the optimal family and parameter for pair-copula $c_{j_e,k_e;D_e}$. This is repeated until all pair-copulas are fitted. Note that it is possible to incorporate a test for independence in the pair-copula selection process \citep[see][]{genest2007everything}. Based on the estimated Kendall's $\tau$ the null hypothesis that the (pseudo) observations come from the independence copula is tested. Then, if the null hypothesis is not rejected for some chosen level $\beta$, the pair-copula to be estimated is chosen to be the independence copula.\\
Regarding the selection of the tree structure, Di\ss mann's algorithm uses a heuristic that models the strongest pairwise dependencies (measured in terms of the absolute empirical Kendall's $\tau$ value) in the lower trees. To be precise, the algorithm starts with the first tree, where all pair-wise empirical Kendall's $\tau$ values are determined and then a maximum spanning tree using the absolute $\tau$ values as weights is selected (e.g.\ by the Algorithm of Prim \citep[][Section 23.2]{thomas2001introduction}). For the next tree, all required pseudo observations are determined as above and for all edges allowed by the proximity condition the Kendall's $\tau$ values are estimated. Again, a maximum spanning tree is selected using these empirical Kendall's $\tau$ values as edge weights, now on the graph constrained by the proximity condition. This procedure is iterated until all $d-1$ trees are selected.\\
\cite{dissmann2013selecting} justify this heuristic by noting that the lowest trees have the greatest influence on the overall fit and thus it is important to model most of the dependence early. Further, they argue that this procedure minimizes estimation errors in higher trees, because typically the dependence decreases when using the algorithm, making rounding errors less severe. As a measure of the strength of dependence Kendall's $\tau$ is used, since it is a rank-based dependence measure facilitating the comparison of different copula families \citep[see e.g.][Ch.\ 5.1.1]{nelsen2007introduction}.\\
Alternative methods using other edge weights for the determination of the maximum spanning tree have been proposed. For example, the fitted pair-copula's AIC or the p-value of a goodness-of-fit test have been used in \cite{czado2013selection}, but they are not commonly used, since all possible pair-copulas have to be estimated first to obtain all the edge weights. This becomes time consuming especially in higher dimensions. Another approach for the tree selection of vines is given by \cite{kurowicka2011optimal}, who also uses the heuristic of modeling most of the dependence in the lowest trees by starting with the last tree, where the edge with the lowest partial correlation is chosen. This is repeated, going backwards, until all trees are specified.\\
So we see that there is already quite a range of structure selection methods. However, all these approaches do not take into account the selected structure's implications on the validity of the simplifying assumption and therefore on the overall model fit. The methods we will propose in \autoref{sec:algorithms} remedy this issue. They are based on a test whether the simplifying assumption is fulfilled for a considered pair-copula term as described in the following section.

\subsection{Test for constant conditional correlations (CCC test)}

Formally, the simplifying assumption requires that all the pair-copulas of the vine appearing in \autoref{eq:vine} can be written as follows:
\[
c_{j_ek_e;D_e}\left(\cdot,\cdot;\ub_{D_e} \right)\equiv c_{j_ek_e;D_e}\left(\cdot,\cdot\right).
\]

This means that the pair-copula is independent of the conditioning variables, implying that the dependence associated with the copula is constant with respect to $\ub_{D_e}$. A stochastic representation of the simplifying assumption is given by $(U_{j_e|D_e} ,U_{k_e|D_e}) \perp \Ub_{D_e}$, denoting that the random variables $U_{j_e|D_e}$ and $U_{k_e|D_e}$ are jointly independent of $\Ub_{D_e}$ \citep{kurz2017pacotest}. Without the simplifying assumption, one would have to allow $c_{j_ek_e;D_e}$ to depend on $\ub_{D_e}$, making it a $(|D_e|+2)$-dimensional function to be estimated (which is $d$-dimensional for the pair-copula in the last tree). This would defeat the whole purpose and convenience of vine models, whose idea it is to express a $d$-dimensional copula just in terms of \textit{bivariate} building blocks. While for low dimensions of the conditioning vector researchers have found ways to relax the simplifying assumption \citep{acar2012beyond,vatter2016generalized,schellhase2016estimating}, in higher dimensions this does not seem feasible. Further discussion and implications of the simplifying assumption can be found in \citep{haff2010simplified,stoeber2013simplified,spanhel2015simplified,killiches2016using}.\\
Recently, \cite{kurz2017pacotest} developed a statistical testing method assessing the severity of the violation of the simplifying assumption for each conditional pair-copula in a vine copula. It tests the null hypothesis that the conditional correlation $\rho_{j_ek_e|D_e}$ associated with pair-copula $C_{j_ek_e;D_e}$ is constant with respect to the conditioning variables $\Ub_{D_e}$. For this purpose, the support $\Omega_0$ of $\Ub_{D_e}$ is divided by a partition $\Gamma:=\{\Omega_1,\ldots,\Omega_L\}$, with $L\in\N$, $\Omega_i\cap\Omega_j=\emptyset$, for $i\neq j$, and $P(\Ub_{D_e}\in \Omega_i) > 0$, for $i=1,\ldots,L$. The idea of the test is that under the null hypothesis the conditional correlation between $U_{j_e|D_e}$ and $U_{k_e|D_e}$ given $\Ub_{D_e}\in \Omega_i$ does not depend on $i$. Hence, denoting $\rho_{\Omega_i} := Corr(U_{j_e|D_e},U_{k_e|D_e}|\Ub_{D_e}\in \Omega_i)$, the null hypothesis is equivalent to testing $\rho_{\Omega_1}=\ldots=\rho_{\Omega_L}$, which we will call the constant conditional correlation (CCC) assumption. To derive a test statistic we observe that for the vector of estimated conditional correlations $\hat R^{(n)}(\Gamma):=(\hat\rho^{(n)}_{\Omega_1},\ldots,\hat\rho^{(n)}_{\Omega_L})'$, estimated by the sample Pearson's correlation coefficient based on an i.i.d.\ sample of size $n$, it holds
\[
\sqrt n(\hat R^{(n)}(\Gamma) - R(\Gamma))\xrightarrow[n\rightarrow\infty]{d} N(\0, \Sigma(\Gamma)),
\]
where $R(\Gamma)$ is the vector of true conditional correlations and $\Sigma(\Gamma)$ is the asymptotic variance-covariance matrix \citep[see][]{kurz2017pacotest}. The asymptotic normality of $\hat R^{(n)}(\Gamma)$ is used to derive an asymptotically $\chi^2$ distributed test statistic by taking differences and normalizing. Defining the matrix $A\in \{-1,0,1\}^{L-1\times L}$ with $A_{ij}=\mathbbm{1}\{i=j\}-\mathbbm{1}\{i=j-1\}$, $i=1,\ldots L-1$, $j=1,\ldots L$, i.e.,
\[
A=\begin{pmatrix}
1 & -1 & & \\
 & 1 & -1 & \\
 &&\ddots &\\
& & 1&-1
\end{pmatrix},
\]
where the omitted entries equal zero, we have that
\[
A\hat R^{(n)}(\Gamma)=\begin{pmatrix}
\hat\rho^{(n)}_{\Omega_1} - \hat\rho^{(n)}_{\Omega_2} \\
\hat\rho^{(n)}_{\Omega_2} - \hat\rho^{(n)}_{\Omega_3} \\
\vdots \\
\hat\rho^{(n)}_{\Omega_{L-1}} - \hat\rho^{(n)}_{\Omega_L}
\end{pmatrix}.
\]
The real-valued test statistic is defined by
\[
T_n(\Gamma) = (A\hat R^{(n)}(\Gamma))' (A\hat \Sigma^{(n)}(\Gamma)A')^{-1}A\hat R^{(n)}(\Gamma),
\]
where $\hat \Sigma^{(n)}(\Gamma)$ is a consistent estimator for $\Sigma(\Gamma)$. Since \cite{kurz2017pacotest} show that under regularity conditions and $H_0$ it holds that
\[
nT_n(\Gamma)\xrightarrow[n\rightarrow\infty]{d}\chi^2_{L - 1},
\]
an asymptotic $\beta$-level test is given by 
\[
\mathbbm{1}\{nT_n(\Gamma)\geq F^{-1}_{\chi^2_{L-1}}(1-\beta)\},
\]
where $F^{-1}_{\chi^2_{L-1}}(1-\beta)$ denotes the $1-\beta$ quantile of the $\chi^2$ distribution with $L-1$ degrees of freedom.\\
Of course, the power of this test highly depends on the chosen partition $\Gamma$. Therefore, \cite{kurz2017pacotest} adjust the test statistic to accommodate for the combination of different partitions and further use decision trees to find partitions where a possible violation of the simplifying assumption is most pronounced. The partition size $L$ is determined by the algorithm and grows with $|D_e|$ \citep[see][for details]{kurz2017pacotest}. Finally, they show that the test has a very high observed power compared to benchmark methods. The test is implemented as function \texttt{pacotest} in the R package \texttt{pacotest} \citep{Rpacotest}.

\section{Two new tree selection algorithms}\label{sec:algorithms}

\subsection{Motivation}\label{sec:uranium}

As a motivation we first consider three-dimensional data sets. Here we know that there are three different possible vine tree structures and that the specification of the first tree already fully describes the whole tree structure. In \autoref{sec:dissmann} we have seen that Di\ss mann's algorithm chooses the tree structure that maximizes the dependence between the variables in the first tree. However, the resulting tree structure might yield a non-simplified vine copula even though the true model is a simplified vine copula. Hence, the model fit of Di\ss mann in terms of log-likelihood might not be optimal.\\
This motivates us to select tree structure by testing which combination of variables is ``most simplified''. In detail, we use the R-function \texttt{pacotest} to test for the hypothesis that the CCC assumption holds for $U_i,U_j|U_k$ with $k\in\{1,2,3\}$ and $\{i,j\}=\{1,2,3\}\backslash k$. Then, we choose the tree structure which has the highest p-value in the second tree.
\paragraph*{Example 1}
We show how this works for a real life data set, namely a subset of the well-known seven-dimensional hydro-geochemical data set first investigated by \cite{cook1986generalized}, consisting of $N=655$ observations of log-concentrations of the three chemicals cobalt ($U_1$), titanium ($U_2$) and scandium ($U_3$) in water samples taken from a river near Grand Junction, Colorado. This data set has been examined with regard to the simplifying assumption by many researchers, e.g.\ \cite{acar2012beyond}, \cite{killiches2016examination} and \cite{killiches2016using}. After transforming the data to the copula scale by applying the empirical probability integral transform, we estimate the Kendall's $\tau$ values for each of the three pairs. They are $\hat\tau_{12}=0.54$, $\hat\tau_{13}=0.36$ and $\hat\tau_{23}=0.44$. Consequently, Di\ss mann's algorithm chooses edges $(1,2)$ and $(2,3)$ corresponding to pair-copulas $c_{12}$ and $c_{23}$ for the first tree resulting in the conditional copula $c_{13;2}$ to be modeled in the second tree. The Di\ss mann algorithm fits all pair-copulas as t-copulas. The log-likelihood and AIC-values of the vine copula fitted by Di\ss mann's algorithm are 428.8 and $-845.6$, respectively (see also Structure 2 in \autoref{tab:uranium3d}). 

\begin{table}[ht]
	\centering
	\small
	\begin{tabular}{ccccccc}
		Structure & Tree 1 & conditional copula & p-value & sum of $\tau$ & log-lik & AIC\\
		\hline
		1 & 2--1--3 & $c_{23;1}$ & 0.29 & $0.54+0.36=0.90$ & 434.9 & $-857.8$\\ 
		2 & 1--2--3 & $c_{13;2}$ & 0.01 & $0.54+0.44=0.98$ & 428.8 & $-845.6$\\
		3 & 1--3--2 & $c_{12;3}$ & 0.19 & $0.36+0.44=0.80$ & 418.5 & $-825.1$\\\Xhline{4\arrayrulewidth}
	\end{tabular}
	\caption{The p-values, sums of Kendall's $\tau$ values, log-likelihoods and AIC-values of the three possible tree structures for the three-dimensional uranium dataset.}
	\label{tab:uranium3d}
\end{table}

Let us now examine the p-values of the test for constant conditional correlations for the three possible tree structures. They are 0.29, 0.01, 0.19 for the conditional copulas $c_{23;1}$, $c_{13;2}$ and $c_{12;3}$, respectively. Hence, we see that the structure chosen by Di\ss mann's algorithm has a strong indication for non-simplifiedness, such that we would rather choose one of the other two structures. The structure corresponding to the highest p-value (Structure 1 in \autoref{tab:uranium3d}) also has a larger sum of Kendall's $\tau$ values in the first tree compared to the other structure with a large p-value ($0.54+0.36=0.90$ vs.\ $0.36+0.44=0.80$), such that we would choose the structure with edges $(1,2)$ and $(1,3)$ in the first and $(2,3;1)$ in the second tree. The fitted vine copula of Structure 1 has a Tawn and a t-copula in the first tree and a BB7 copula in the second tree. With a log-likelihood of 434.9 and an AIC of $-857.8$ it provides a better model fit than the one selected by Di\ss mann's algorithm. For completeness we note that Structure 3 in \autoref{tab:uranium3d} has the worst fit, with the lowest sum of $\tau$ in the first tree and rather low p-value for the conditional copula in the second tree. All in all, with Structure 1 we found a way of describing this data set using a simplified vine copula, whereas in the recent literature it seemed necessary to use a non-simplified vine copula model when Structure 2 was used.\\

Extending this idea to higher dimensions, things get more complicated because there are superexponentially more tree structures to choose from. While e.g.\ in five dimensions there are still ``only'' 480 possible tree structures, in ten dimensions they already explode to $4.8705\cdot 10^{14}$. See \cite{morales2011count} for a general formula for counting the number of different vine structures. The question is now how to incorporate the information about the tests concerning the simplifiedness of the conditional copulas for the determination of the best vine tree. 
\subsection{Algorithm 1: Regular vine structure selection using CCC test based weights}\label{sec:alg1}
The first approach that we propose is to choose the first tree (where there are no explicit considerations about simplifiedness yet) according to Di\ss mann's approach and then to incorporate the CCC tests in the higher level trees. In each tree, the test's p-values of all edges allowed by the proximity condition are calculated and used to find the maximum spanning tree. This would constitute a compromise between both approaches reflecting as much of the dependence as possible in the first tree and accounting for non-simplifiedness in the trees where conditional copulas have to be fitted. Further, one can also assign each edge a score combining both the edge's p-value and estimated absolute Kendall's $\tau$. Hence, the score of an edge $e$ would be defined as
\[
s_{\alpha}(e):=\alpha \cdot f_p(e) + (1-\alpha)\cdot f_{\tau}(e),
\]
where $\alpha$ is a weighting factor and the functions $f_p$ and $f_{\tau}$ map each edge $e$ allowed by the proximity condition to a score regarding its degree of simplifiedness and absolute Kendall's $\tau$ values, respectively. For example, $f_p$ could be the function mapping each edge to its associated rank of the p-value arising from the CCC test, i.e.\ $f_p(e_i)=i$, if the edges $e_i$ are ordered such that $\text{p-value}(e_1)<\text{p-value}(e_2)<\ldots<\text{p-value}(e_N)$. Here, $N$ is the number of edges allowed by the proximity condition. Similarly, $f_{\tau}$ could rank the edges by their absolute empirical Kendall's $\tau$ values, such that the scales of $f_p$ and $f_{\tau}$ would coincide. The weighting factor $\alpha$ can be chosen by the user. We will see in \autoref{sec:choice_alpha} that values around 0.6 yield good fits in terms of the AIC. After having determined the score of each edge, a maximum spanning tree on the edges using the scores $s_{\alpha}(e)$ is constructed. Then, similar to Di\ss mann's algorithm the pair-copulas corresponding to the chosen tree are fitted (with or without an independence test). \autoref{Alg1} implements this procedure.\\

\begin{algorithm}
	\caption{Regular vine structure selection using CCC test based weights}
	\label{Alg1}
	\textbf{Input:} $d$-dimensional copula data, weighting factor $\alpha$, edge score functions $f_p$ and $f_{\tau}$.
	\begin{algorithmic}[1]
		\State Tree 1: estimate absolute empirical Kendall's $\tau$ values between all variables and find the corresponding maximum spanning tree. For each selected edge of the tree, fit a pair-copula based on the copula data and determine the corresponding pseudo observations for the next tree.
		\For {$m=2,\ldots,d-1$}
		\State Tree $m$: for all edges $e$ allowed by the proximity condition, calculate the score $s_{\alpha}(e)=\alpha \cdot f_p(e) + (1-\alpha)\cdot f_{\tau}(e)$ using the pseudo observations and find the corresponding maximum spanning tree. For each selected edge of the tree, fit a pair-copula based on the pseudo observations and determine the corresponding pseudo observations for the next tree.
		\EndFor
	\end{algorithmic}
	\textbf{Output:} R-vine with tree structure focused on constant conditional dependencies as well as large dependencies in lower trees.
\end{algorithm}
		
\paragraph*{Example 2}

We show how the algorithm works in detail for a 5-dimensional example. At first we generate a random R-vine object with random tree structure, families and parameters and simulate a sample of size 1000 from it (exact details on how the random vine is generated can be found in \autoref{sec:simstudy}). The resulting true vine specification is a D-vine with pair-copulas given in \autoref{tab:exa}.
\begin{table}[ht]
	\centering
	\small
	\begin{tabular}{l|l}
		Tree 1 & Tree 2\\
		\hline
			\parbox[t]{7cm}{$C_{23}$: Clayton (0.44) ($\tau_{23} = 0.18$)\\
			$C_{24}$: type 1 Tawn (2.43, 0.81) ($\tau_{24} = 0.51$)\\
			$C_{15}$: type 2 Tawn (6, 0.43) ($\tau_{15} = 0.39$)\\
			$C_{45}$: survival Gumbel (2.69) ($\tau_{45} = 0.63$)}
		&
			\parbox[t]{8cm}{$C_{34;2}$: survival BB6 (1.07,2.31) ($\tau_{34;2} = 0.58$)\\
			$C_{25;4}$: BB8 (1.62, 0.7) ($\tau_{25;4} = 0.1$) \\
			$C_{14;5}$: 270 degree rotated type 1 Tawn ($-3.89$,\\ \mbox{}\quad\quad\quad 0.54) ($\tau_{14;5} = -0.45$)}\\
		\hline
		Tree 3 & Tree 4\\
		\hline
		\parbox[t]{7cm}{$C_{35;24}$: survival Joe (2.31) ($\tau_{35;24} = 0.42$)\\
		$C_{12;45}$: Frank (18.67) ($\tau_{12;45} = 0.8$)}
		&
		\parbox[t]{8cm}{$C_{13;245}$: survival BB1 (0, 2.39) ($\tau_{13;245} = 0.58$)}\\
		\Xhline{4\arrayrulewidth}		
	\end{tabular}
	\caption{D-vine copula specification of Example 2.}
	\label{tab:exa}
\end{table}

\begin{figure}[!htb]
	\centering
	\includegraphics[trim=0cm 0cm 0cm 0cm,clip,width=0.7\textwidth]{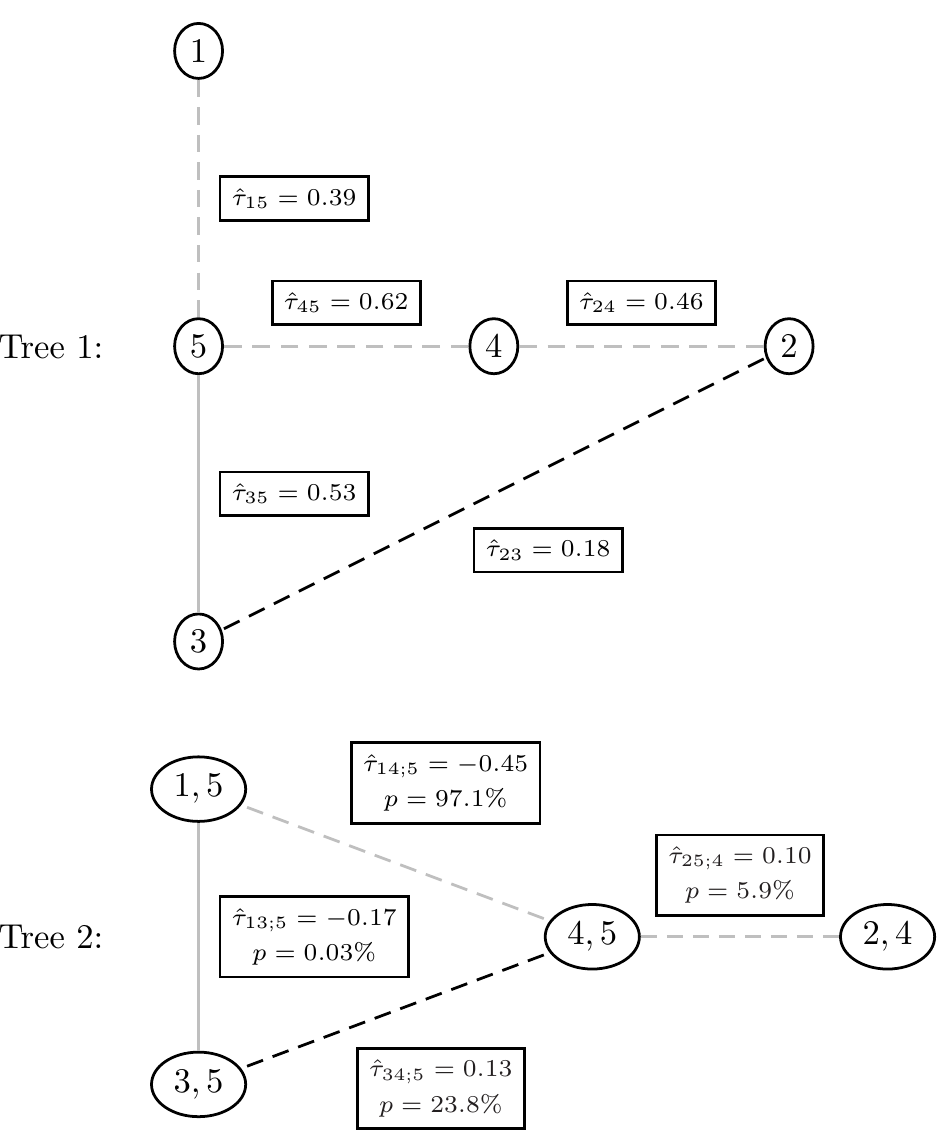}
	\caption{First two trees selected by Di\ss mann's algorithm (gray). In Tree 1, the empirical Kendall's $\tau$ values of each edge are given in the boxes and the dashed lines correspond to the true tree structure. In Tree 2, additionally to Kendall's $\tau$, the p-values of the conditional correlation test are given and the dashed lines mark the tree selected by \autoref{Alg1}.}
	\label{fig:Example}
\end{figure}

The chosen vine's first tree connects the edges $(2,3)$, $(2,4)$, $(1,5)$ and $(4,5)$. In a first step we fit an R-vine to the simulated data using the function \texttt{RVineStructureSelect} in which Di\ss mann's algorithm is implemented. The resulting first two trees are displayed in \autoref{fig:Example} as the gray graphs. 

The first tree of the fitted R-vine is quite close to the true one (highlighted by a dashed line). Edges $(2,4)$, $(1,5)$ and $(4,5)$ are selected and the true copula families are chosen with parameters very close to the true ones ($\hat\tau_{24}=0.46$, $\hat\tau_{15}=0.39$ and $\hat\tau_{45}=0.62$). Only edge $(3,5)$ is selected instead of $(2,3)$ since it has a higher empirical Kendall's $\tau$ (0.53 compared to 0.18).\\
Due to the proximity condition in the second tree, node $(2,4)$ can only be connected to $(4,5)$ (since it is the only other node containing a 4 or a 2), resulting in the edge $(2,5;4)$. Regarding nodes $(1,5)$, $(3,5)$ and $(4,5)$ the proximity condition allows all of these nodes to be connected. Only focusing on the absolute empirical Kendall's $\tau$ value, Di\ss mann's algorithm chooses edges $(1,3;5)$ and $(1,4;5)$, since $|\hat\tau_{14;5}|>|\hat\tau_{13;5}|>|\hat\tau_{34;5}|$. Altogether, the selected second tree constitutes a path and therefore the higher level trees are already determined by the proximity condition. The AIC of the R-vine fitted by Di\ss mann's algorithm is $-7775.5$.
		
Next, we consider the results of our proposed algorithm on this data set. Of course, the selected first tree coincides with the one from Di\ss mann's algorithm. However, in the second tree, when choosing which of the nodes $(1,5)$, $(3,5)$ and $(4,5)$ to connect, we see that even though edge $(1,3;5)$ has a slightly larger absolute Kendall's $\hat\tau$ than edge $(3,4;5)$, its p-value ($0.03\%$) is much smaller than the $23.8\%$ of edge $(3,4;5)$. Thus, for $\alpha\geq 0.5$ our proposed algorithm would choose edges $(3,4;5)$ and $(1,4;5)$ for the second tree (cf.\ the dashed graph in the lower panel of \autoref{fig:Example}). After having fit the third and fourth trees accordingly, the overall resulting R-vine proves to yield a much better fit than Di\ss mann's algorithm with an AIC of $-8272.9$.

In the simulation study presented in \autoref{sec:simstudy} we repeat this procedure 1000 times and find that \autoref{Alg1} achieves a better or equal AIC-value in 80.5 percent of the simulations. 

\subsection{Algorithm 2: C-vine structure selection using CCC test based weights}

One major disadvantage of \autoref{Alg1} is that the overall tree structure is already strongly affected by the first tree due to the proximity condition. For example, if the first tree is fitted with a path-like structure, then the whole vine is automatically a D-vine since the proximity condition only allows for one possible (also path-like) structure in the higher trees. In this case we cannot account for a possible non-simplifiedness of the model at all. For example, in three dimensions \autoref{Alg1} would always yield the same structure as the Di\ss mann algorithm since for three nodes every tree constitutes a path. So, e.g.\ for the three-dimensional uranium example described in \autoref{sec:uranium} where the consideration of the test for the simplifying assumption improved the structure fit, the proposed algorithm would be futile.

Thus we have to find a way how to sequentially select a vine structure incorporating in the current tree the information about which conditional correlations are not constant in the subsequent tree while still keeping the subsequent graph as flexible as possible. We have seen that a path-like structure in the first tree already determines the structure of the remaining trees. Conversely, we know that we retain the highest flexibility by fitting a star-like structure in the first tree since then the proximity condition imposes no restriction (such that all nodes in the subsequent tree are allowed to be linked by an edge). Following this logic for all trees, a star-like structure is reasonable for every tree level. The root node $v_m$ of each star should be chosen that on the one hand the sum of absolute empirical Kendall's $\tau$ values between this root node and all the other variables (conditioned on the earlier root nodes $v_1,\ldots,v_{m-1}$) is large. On the other hand, conditioned on the root node $v_m$ (and all the earlier root nodes) pairs of the remaining variables should have rather constant conditional correlations, i.e.\ the p-values of the CCC test conditioning on the root nodes should be large.\\
Formally, at tree level $m$ we assign each remaining node $v\in N_m:= \{1,\ldots,d\}\backslash\{v_1,\ldots,v_{m-1}\}$ the score $s_{\alpha}(v)$ (operating on nodes $v$ instead of edges $e$ as in \autoref{Alg1}) defined as
\[
s_{\alpha}(v) = \alpha \cdot g_p(v) + (1-\alpha)\cdot g_{\tau}(v).
\]
We then choose the one with the highest score as a root node for tree $m$. Again, $\alpha$ is a weighting factor and $g_p$ and $g_{\tau}$ are functions mapping a node to ranked scores regarding the CCC tests and the Kendall's $\tau$ values, respectively. In particular, $g_p$ calculates for all $v\in N_m$ the p-value score
\[
p(v):=\sum_{i,j\in N_m\backslash v,\, i\neq j} r(p_{ij;v_1,\ldots,v_{m-1},v}),
\]
and then maps each $v$ to its rank among all p-value scores $\{p(v)\}_{v\in N_m}$ (e.g.\ the root node with the smallest p-value score would be assigned rank 1). Here, $p_{ij;v_1,\ldots,v_{m-1},v}$ denotes the p-value of the test for constant conditional correlation of $C_{ij;v_1,\ldots,v_{m-1},v}$ and the function $r$ maps a p-value $p_{i_0j_0;v_1,\ldots,v_{m-1},v_0}$ to its rank among all possible p-values $p_{ij;v_1,\ldots,v_{m-1},v}$ with $i,j,v\in N_m$ pairwise distinct. Instead of choosing the function $r$ to be the rank transformation, other transformations such as the logarithm are possible. However, the results of the simulation study show that the rank transformation yields the best results (see \autoref{sec:choice_r}).\\
Regarding Kendall's $\tau$, $g_{\tau}$ similarly calculates the $\tau$ scores
\[
t(v):=\sum_{i\in N_m\backslash v} |\hat{\tau}_{iv;v_1,\ldots,v_{m-1}}|,\, v\in N_m 
\]
and then assigns each $v$ to its corresponding rank among these $\tau$ scores. The Kendall's $\tau$ associated to $c_{iv;v_1,\ldots,v_{m-1}}$ is estimated as the empirical Kendall's $\tau$ between the pseudo observations $\hat\ub_{i|v_1,\ldots,v_{m-1}}$ and $\hat\ub_{v|v_1,\ldots,v_{m-1}}$ and is denoted by $\hat\tau_{iv;v_1,\ldots,v_{m-1}}$.\\
After the optimal root node is found, similar to the other algorithms all pair-copulas implied by the resulting tree have to be fitted (with or without independence test) for the calculation of the pseudo observations of the next tree level (see \autoref{Alg2}).

\begin{algorithm}
	\caption{C-vine structure selection using CCC test based weights}
	\label{Alg2}
	\textbf{Input:} $d$-dimensional copula data, weighting factor $\alpha$, node score functions $g_p$ and $g_{\tau}$.
	\begin{algorithmic}[1]
		\For {$m=1,\ldots,d-1$}
		\State Tree $m$: Determine the optimal root node by evaluating score $s_{\alpha}(v) = \alpha \cdot g_p(v) + (1-\alpha)\cdot g_{\tau}(v)$ for all nodes $v\in N_m$ and choose the maximal one. Fit pair-copulas between this root node and the remaining nodes based on the pseudo observations. Calculate the corresponding pseudo observations needed for the tree selection in the next step.
		\EndFor
	\end{algorithmic}
	\textbf{Output:} C-vine with tree structure focused on constant conditional dependencies as well as large dependencies in lower trees.
\end{algorithm}

Using this algorithm with any $\alpha>0.5$, we retrieve the AIC-optimal tree structure for the uranium example discussed in \autoref{sec:uranium}. 
\paragraph*{Example 2 (continued)}
We revisit the 5-dimensional example from the previous section and want to apply \autoref{Alg2} to the simulated data (with $\alpha=0.6$ and rank-transformed p-values). \autoref{tab:Ex2ctd} displays the information needed to find the optimal root node $v_1$ in the first tree of the C-vine.

\begin{table}[ht]
	\centering
	\small
	\begin{tabular}{c|cc|cc|c}
		node $v$ & $p(v)$ & $\rank(p(v))$ & $t(v)$ & $\rank(t(v))$ & $s_{\alpha}(v)$ \\
		\hline
		1 & 91 & 3 & 1.04 & 1 & 2.2 \\ 
		2 & 89 & 2 & 1.47 & 3 & 2.4 \\ 
		3 & 105 & 4 & 1.31 & 2 & 3.2 \\ 
		4 & 69 & 1 & 1.72 & 4 & 2.2 \\ 
		5 & 111 & 5 & 1.98 & 5 & 5.0 \\ \Xhline{4\arrayrulewidth}
	\end{tabular}
	\caption{p-value scores $p(v)$ and $\tau$ scores $t(v)$ with associated ranks, and the overall score $s_{\alpha}(v)$ with $\alpha=0.6$ for the determination of the root node $v_1$ in the first tree of the C-vine fitted by \autoref{Alg2}.}
	\label{tab:Ex2ctd}
\end{table}

For example, the p-value score $p(1)=91$ is the sum of the ranks of the 6 p-values $p_{ij;1}$, $i,j\in\{2,3,4,5\}$, $i\neq j$, among the 30 possible p-values. Since 91 is the third largest p-value score, its rank among the p-value scores is 3. Similarly, since $\sum_{i=2}^5 |\hat{\tau}_{i1}|=1.04$ is the smallest of the $\tau$ scores node 1 gets assigned rank 1. Hence, the overall score is calculated as $s_{\alpha}(v)=0.6\cdot 3 + 0.4\cdot 1 = 2.2$. Doing this for the other possible root nodes we see that node 5 has the largest p-value and $\tau$ scores and therefore with an overall score of $5.0$ is selected as the first root node. This procedure is repeated for all trees yielding a C-vine with root node ordering 5--4--2--1--3 and an AIC of $-7446.6$. While we see that this model has a higher AIC than the ones selected by \autoref{Alg1} and by Di\ss mann's algorithm (which may be due to the fact that the true model is a D-vine and \autoref{Alg2} fits a C-vine), it still finds the AIC-optimal out of the 60 possible C-Vines structures. Further, we will see in \autoref{sec:simstudy}, where this procedure is repeated 1000 times, that in 73.6\% of the time \autoref{Alg2} finds a C-vine with a better AIC than the one of Di\ss mann's chosen R-vine.

\section{Simulation study}\label{sec:simstudy}

\subsection{Setup}\label{sec:setup}
We perform an extensive simulation study, evaluating the performance of the two proposed algorithms in many different scenarios. For each scenario, we repeat $R=1000$ simulations of sample size $n$ of randomly generated $d$-dimensional R-vine copulas. For these, we sample uniformly one out of the $2^{(d-2)(d-3)/2}$ different structure matrices with natural ordering as described in \cite{joe2011regular} using the R function \texttt{RVineMatrixSample} from \texttt{VineCopula}; for each pair-copula, a random copula family out of the 12 families currently implemented in the package \texttt{VineCopula} is selected (see \autoref{app:families} for a list of the implemented families). Regarding the copulas' parameters, $Beta(2,2)$ distributed Kendall's $\tau$ values are generated and multiplied by $-1$ with a probability of 50\% (note that the densities of some families have to be rotated by 90 degrees to accommodate negative dependence). The $Beta(2,2)$ distribution is symmetric around its mode 0.5 with a variance of 0.05 and has a 95\% confidence interval given by $[0.09,0.91]$. Next, for the two-parametric copula families the second parameter is randomly generated from suitable distributions (see \autoref{app:secondpar} for details). Finally, the first parameter of all pair-copulas is derived from the sampled Kendall's $\tau$ values and the second parameters (where required) using R function \texttt{BiCopTau2Par} \citep[see][for details]{joe1997multivariate}. An exemplary random 5-dimensional R-vine is given in the \autoref{tab:exa} of \autoref{sec:alg1}.\\
In the simulation study, for each of the samples of size $n$ from these random R-vines we will calculate the AIC-values of the vines fitted using our proposed algorithms (here without the mentioned independence tests). The AIC-values will be compared to those of the fitted Di\ss mann vines. Before presenting the results of the simulation study, we shortly investigate the influence of the weighting factor $\alpha$ on the AIC-values of the models fitted by our algorithms. 

\subsection{Choice of weighting factor $\alpha$}\label{sec:choice_alpha}

\begin{figure}[!htb]
	\centering
	\includegraphics[trim=0cm 0cm 0cm 0cm,clip,width=0.32\textwidth]{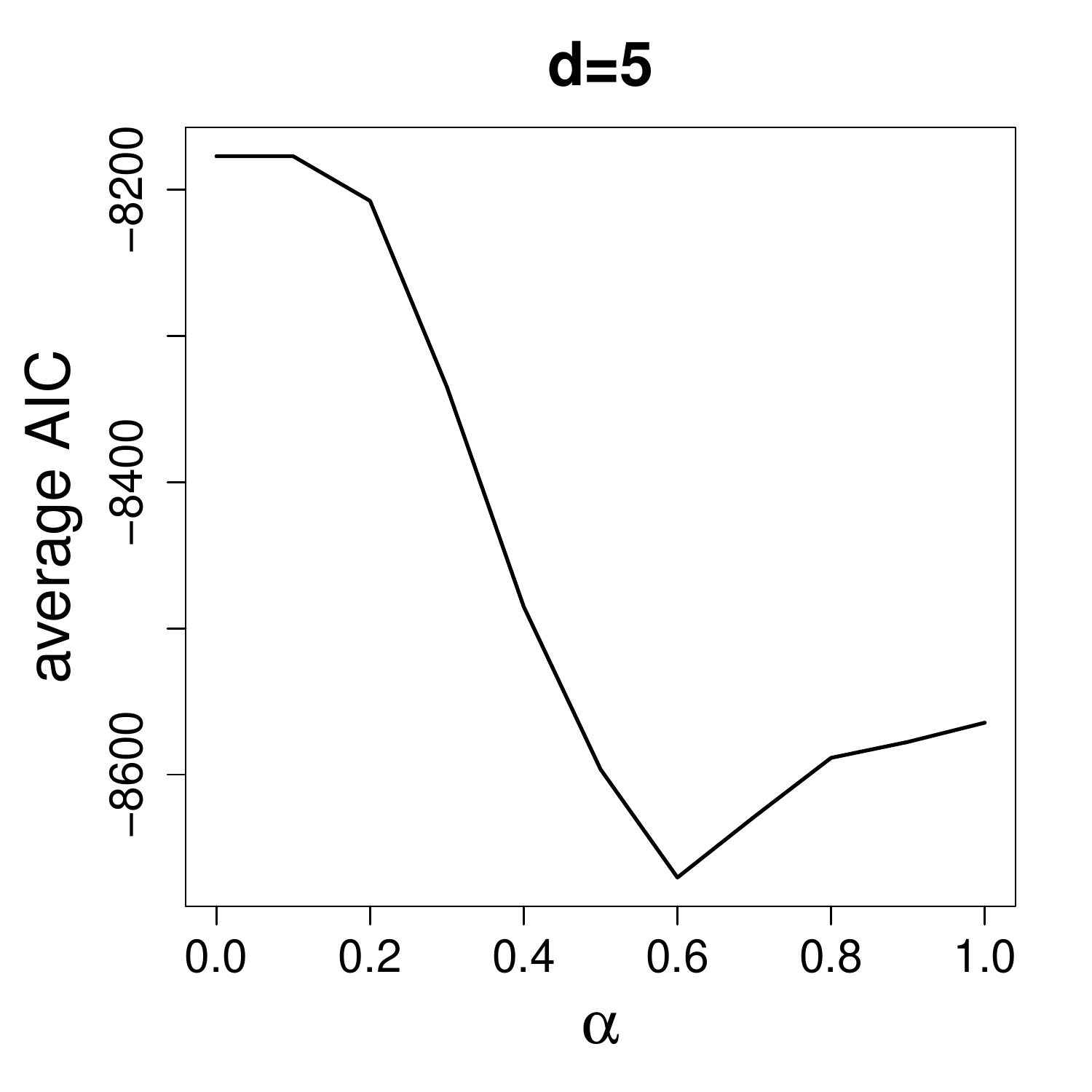}
	\includegraphics[trim=0cm 0cm 0cm 0cm,clip,width=0.32\textwidth]{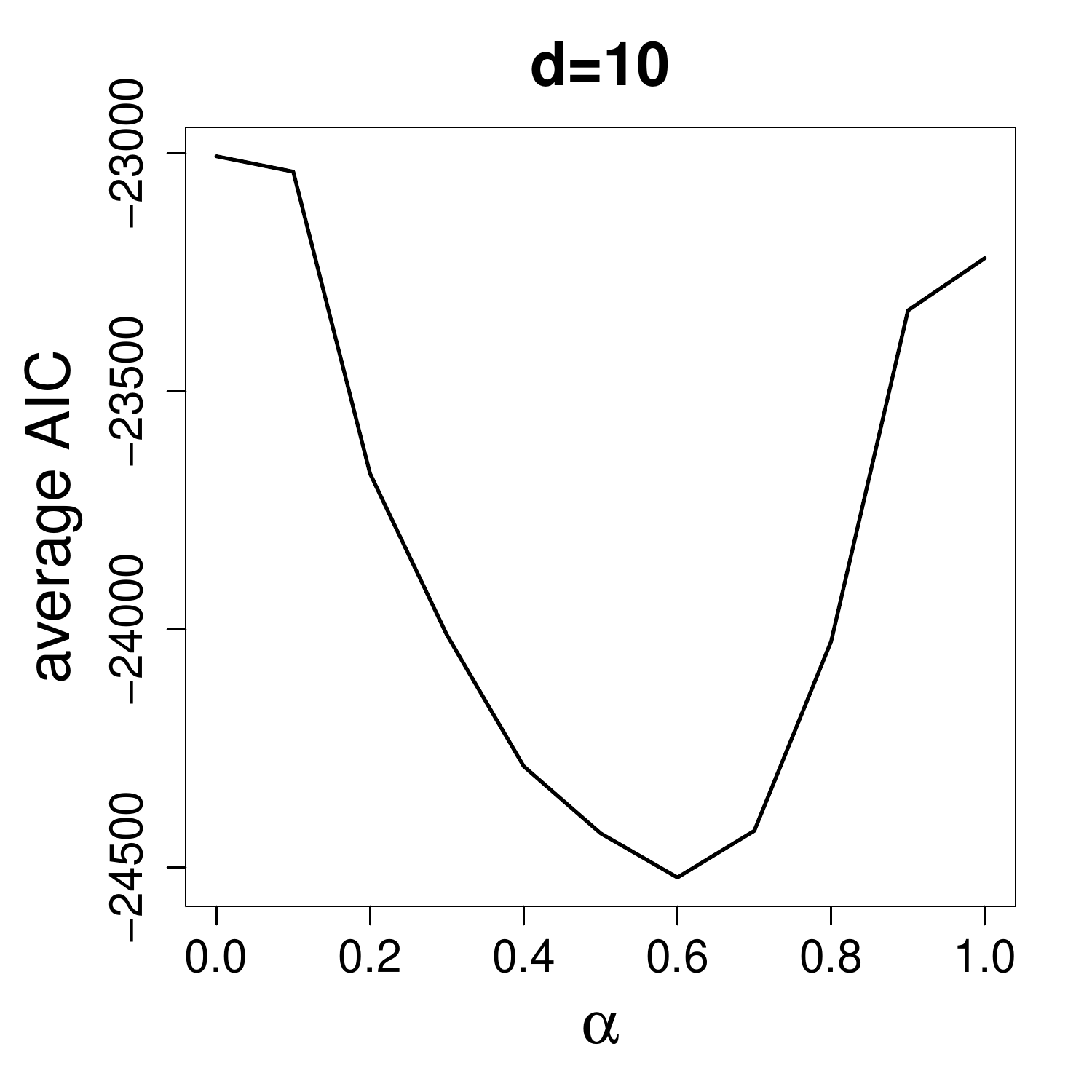}
	\includegraphics[trim=0cm 0cm 0cm 0cm,clip,width=0.32\textwidth]{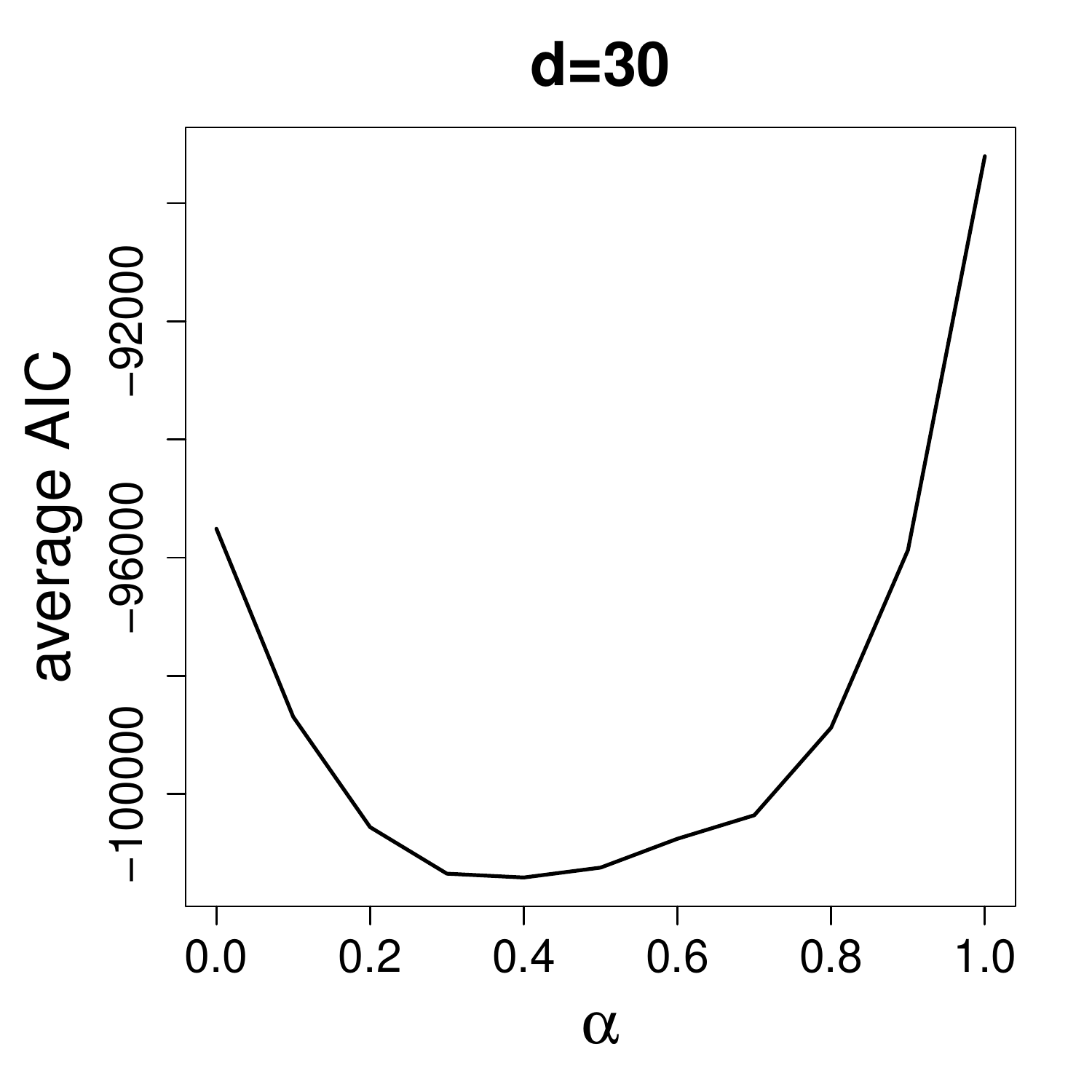}
	\caption{AIC-values of vine copulas fitted by \autoref{Alg2} depending on the weighting factor $\alpha$ in the setting $n=1000$ and $d=5,10,30$, averaged over 1000 repetitions.}
	\label{fig:choiceofalpha}
\end{figure}

In the simulation study we will consistently use the weighting factor $\alpha=0.6$. This choice is justified because we heuristically observe that the AIC of our fitted values on average is lowest for medium sized weighting factors with the best performances for $\alpha=0.6$. 
We generate 1000 random $d$-dimensional vine copulas (as described in \autoref{sec:setup}), simulate a sample of size $n$ from each copula and apply our algorithms to the sample using weighting factors $\alpha\in\{0,0.1,0.2,\ldots,0.9,1\}$. For each $\alpha$ we compute the AIC of the fitted models averaged over the 1000 repetitions. Overall, we find that in most scenarios $\alpha=0.6$ yields the lowest average AIC or close to it. \autoref{fig:choiceofalpha} displays exemplary results for \autoref{Alg2} in the setting $n=1000$ and $d=5,10,30$.

We observe a U-shape of the plots indicating that choosing the weighting factor too small or too large does not yield optimal fits. So, focusing exclusively on the Kendall's $\tau$ values or the p-values of the CCC test apparently is not sufficient. For $d=5$ and $d=10$ the minimum of the average AIC-values is attained for $\alpha=0.6$ and in dimension thirty for $\alpha=0.4$ with the average AIC of $\alpha=0.6$ being still quite close to the minimum. The results for sample sizes $n=400$ and $n=3000$ given in \autoref{app:choiceofalpha} are very similar and lead the same conclusion of $\alpha=0.6$ being optimal. Finally, the same study applied for \autoref{Alg1} also yields an optimum of $\alpha=0.6$.  

\subsection{Results}
In our simulation study we let the dimension $d$ of the randomly generated vine copulas vary between 3 and 50 and the sizes of the simulated samples $n$ between 400 and 3000. The percentage of times where our algorithms performs better or equal than Di\ss mann is given in \autoref{tab:simstudy} (the percentages of equal performance are given in brackets and are omitted if they are 0). The results of the table are also visualized in \autoref{fig:simstudy_results} of \autoref{sec:plot_simstudy}.

\begin{table}[ht]
	\centering
	\scriptsize
	\begin{tabular}{c|c|lllllllll}
		&&\multicolumn{9}{c}{$d$}\\
		\hline
		Algorithm & $n$ & 3 & 4 & 5 & 7 & 10 & 15 & 20 & 30 & 50 \\ 
		\hline
		\multirow{3}{*}{Algorithm 1} & $400$ & 100 (100) & 92.2 (89.0) & 80.8 (69.3) & 54.9 (30.8) & 41.9 (5.9) & 34.8  (0.2)& 30.0 & 26.5 & 22.1 \\
		& $1000$ & 100 (100) & 93.1 (89.0) & 82.3 (69.7) & 59.9 (32.3) & 41.4 (4.8) & 35.9 (0.1) & 30.3 & 28.9 & 27.3 \\
		& $3000$ & 100 (100) & 92.6 (88.6) & 79.9 (68.0) & 58.3 (31.3) & 41.6 (4.2) & 36.2 (0.2) & 35.3 & 21.8 & 18.2 \\
		\hline
		\multirow{3}{*}{Algorithm 2} & $400$ & 84.4 (49.6) & 68.8 (3.1) & 67.7 & 75.9 &   82.0 & 93.2 & 96.7 & 99.2 & 99.8 \\ 
		& $1000$ & 91.1 (51.5) & 75.5 (3.8) & 73.6 & 80.2 & 88.7 & 95.8 & 99.0 & 99.6 & 100 \\ 
		& $3000$ & 93.4 (47.6) & 77.1 (3.7) & 75.8 & 82.2 & 91.7 & 96.8 & 99.2 & 100 & 100 \\\Xhline{4\arrayrulewidth}
	\end{tabular}
	\caption{Percentages of better or equal performance regarding the AIC-value of the two algorithms compared to Di\ss mann's algorithm for different dimensions $d$ and sample sizes $n$ based on 1000 data sets sampled from randomly generated R-vines (in brackets the percentages of equal performance are given).}
	\label{tab:simstudy}
\end{table}

We see that in general our two proposed algorithms perform very well compared to Di\ss mann's algorithm. Of course, in low dimensions \autoref{Alg1} performs very similar to Di\ss mann's algorithm since there the first tree greatly determines the model fit. So as we already noted, in three dimensions \autoref{Alg1} and Di\ss mann's algorithm always find the same vine structure and even in four and five dimensions they coincide in almost 90\% and 70\% of the times, respectively. In higher dimensions, when the selected tree structures differ more often, \autoref{Alg1} manages to find a better structure than Di\ss mann's algorithm in more than one third of the simulations for dimensions 10-15 and around one fourth of the time for dimensions larger than 20. Further, it seems that the sample size has no effect on which of the two algorithms performs better, since the percentages stay rather constant when the sample size changes from 400 to 3000.\\
Concerning \autoref{Alg2} the results are even more promising. In every scenario it is better or equal than Di\ss mann's algorithm in more than two thirds of the time. Especially in high dimensions ($d\geq 15$) it outperforms in more than 90\% of simulations. Moreover, we note that a larger sample size helps to increase the advantage of \autoref{Alg2} even further with increasing percentages for $n$ going from 400 to 3000, regardless of the considered dimension.\\

\begin{figure}[!htb]
	\centering
	\includegraphics[trim=0cm 1.5cm 0cm 1.7cm,clip,width=1\textwidth]{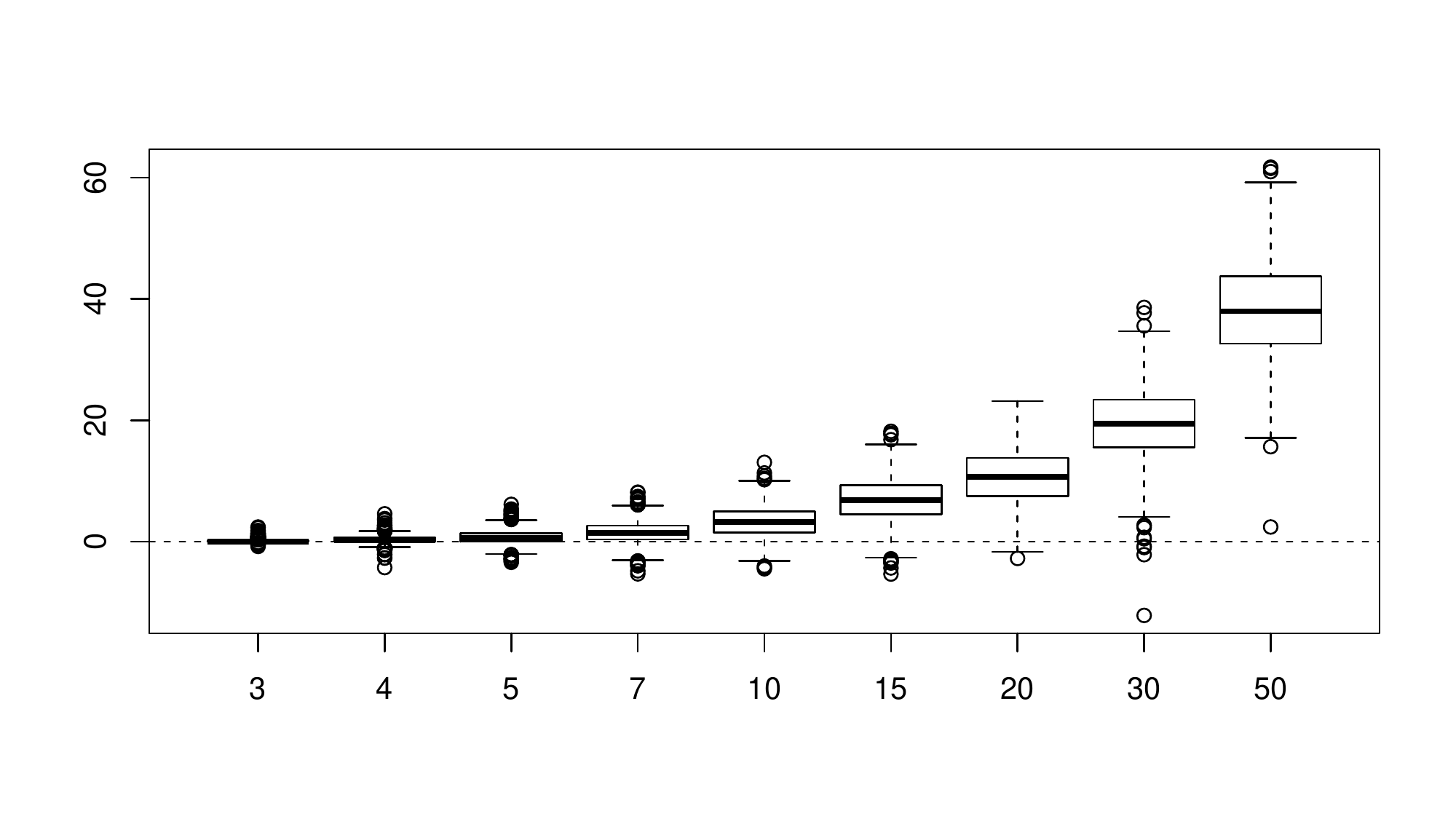}
	\caption{Boxplots of the difference between the AIC-values per observation of the vines chosen by Di\ss mann's algorithm and \autoref{Alg2}.}
	\label{fig:AIC_boxplot}
\end{figure}

Regarding the question, how much better \autoref{Alg2} performs than Di\ss mann's algorithm we present in \autoref{fig:AIC_boxplot} boxplots of the difference between the AIC-values per observation of the models found by Di\ss mann's and our second algorithm for $n=1000$. Positive values imply a worse performance of the Di\ss mann algorithm.

The boxplots affirm the results from \autoref{tab:simstudy}. The differences between the AIC of Di\ss mann's algorithm and \autoref{Alg2} are mostly positive and become larger as the dimension increases. We further see that the magnitude of AIC improvements is much larger than that of the AIC losses. Since the values are shown per observations, we observe AIC improvements of up to 60,000 in 50 dimensions. Due to the scaling the values in the lower dimensions appear to be close to zero. However, we performed t-tests with the null hypothesis that the mean of the AIC differences is zero and could reject it in favor of \autoref{Alg2} with p-values smaller than $10^{-30}$ for all dimensions. The boxplots in the scenarios $n=400$ and $n=3000$ are very similar and can be seen in \autoref{app:boxplots}.\\
In our simulation study the pair copula's Kendall's $\tau$ values were sampled from $[-1,1]$. Since many real life data sets (especially in finance) consist of variables that are pairwise positively dependent we repeated our simulation study sampling only positive Kendall's $\tau$ values. An excerpt of the results is given in the first two rows of \autoref{tab:simstudy_pos_ind} in \autoref{app:posdep_ind}. The results are quite similar with better performances in low  and worse performance in high dimensions.\\
In the construction of the algorithms we noted that the pair-copula selection can be done with or without a prior independence test deciding whether the independence copula can be used. The results above were obtained without the use of independence test. In the last two lines of \autoref{tab:simstudy_pos_ind} in \autoref{app:posdep_ind} we present the results in the positive dependence case when our as well as Di\ss mann's algorithms use the independence test at level $\beta=0.05$. All the results are very similar to those without the independence test implying that the relative performances of the algorithms do not strongly depend on the use of the independence test.\\
Finally, for both algorithms we chose the rank functions for scoring the Kendall's $\tau$ and p-values of the test for constant conditional correlations. This is justified in \autoref{sec:choice_r}.

\subsection{Computational times and a faster version of \autoref{Alg2}}
We consider whether the better performance of our algorithm comes at a higher computational cost. The average computational times (based on 100 repetitions) of fitting a vine copula with the three competing algorithms depending on the dimension $d$ are displayed for $n=1000$ in \autoref{tab:runtime}.

\begin{table}[ht]
	\centering
	\begin{tabular}{l|rrrrrrrrr}
		&\multicolumn{9}{c}{$d$}\\
		\hline
		Algorithm & 3 & 4 & 5 & 7 & 10 & 15 & 20 & 30 & 50 \\
		\hline
		Di\ss mann  & 0.42 & 0.92 & 1.47 & 3.14 & 6.63  & 15.6 & 24.6  & 67.8 & 155.3 \\ 
		Algorithm 1 & 0.41 & 0.83 & 1.42 & 2.94 & 6.42  & 15.2 & 23.5 & 65.1  & 157.7 \\ 
		Algorithm 2 & 0.49 & 1.20 & 2.37 & 6.65 & 20.08 & 72.9 & 154.6 & 731.2 & 3409.4 \\\Xhline{4\arrayrulewidth}
	\end{tabular}
	\caption{Average computational times of the three algorithms for $n=1000$ in different dimensions.}
	\label{tab:runtime}
\end{table}

We see that the computational times of Di\ss mann's algorithm and \autoref{Alg1} are always very close to each other with \autoref{Alg1} being even faster most of the times. \autoref{Alg2} clearly is much slower than the other algorithms since for the determination of the optimal root node each possible pair-copula of the current tree level has to be fitted for the calculation of all the required p-values. This effect increases with the dimension making \autoref{Alg2} comparably fast in low dimensions, only 3 times slower in 10 dimensions, but roughly 20 times slower for $d=50$.\\
A possible way to decrease the computational time of \autoref{Alg2} is given by the following adjustment: in a first step we restrict the allowed pair-copula families exclusively to the Gaussian copula and proceed as before to determine the estimated tree structure. In a second step all pair-copulas appearing in this tree structure are fitted again, now allowing for all pair-copula families. The idea is that the resulting vine copula models hopefully do not differ too much from the unadjusted fit, since the p-values of the CCC tests should not be very sensitive to the chosen families used to determine the pseudo observations and thus yield similar choices for the optimal root nodes. At the same time the computational effort is greatly reduced, since the most time consuming step of \autoref{Alg2} is the fitting of the pair-copulas to determine the p-values of the CCC tests. \autoref{tab:Alg2_adj} displays the percentages of computational times and AIC-values of the adjusted version of \autoref{Alg2} relative to the unadjusted version for dimensions $d=5,10,30$ and sample sizes $n=400,1000,3000$.

\begin{table}[ht]
	\centering
	\begin{tabular}{r|ccc|ccc|ccc}
		$d$ & \multicolumn{3}{c|}{$5$} & \multicolumn{3}{c|}{$10$} & \multicolumn{3}{c}{$30$}\\
		\hline
		$n$ & 400 & 1000 & 3000 & 400 & 1000 & 3000 & 400 & 1000 & 3000 \\
		\hline
		\% time  & 52.0 & 52.1 & 51.9 & 31.2 & 29.6  & 28.2 & 15.8  & 14.0 & 12.9 \\ 
		\% AIC & 98.5 & 97.7 & 97.1 & 98.4 & 98.7  & 97.4 & 99.0 & 98.2  & 98.2 \\ 
	\Xhline{4\arrayrulewidth}
	\end{tabular}
	\caption{Percentages of computational times and AIC-values of the adjusted version of \autoref{Alg2} relative to the unadjusted version for dimensions $d=5,10,30$ and sample sizes $n=400,1000,3000$.}
	\label{tab:Alg2_adj}
\end{table}

We see that the adjustment greatly reduces the computational time making the algorithm almost as fast as Di\ss mann's algorithm. Further, the loss in performance is negligibly small with relative performances ranging between 97 and 99 percent. Thus, we found a way to significantly improve the computational times of \autoref{Alg2} without losing too much performance.

\section{Real data examples}\label{sec:realdata}

In order to assess the performance of our proposed algorithms when confronted with real data, we revisit several data sets that have been used in the recent vine copula literature. For the description of each of the data sets we refer the reader to the respective references. The uranium data set, of which we already examined a three-dimensional subset in \autoref{sec:uranium} was introduced in \cite{cook1986generalized}. The bike data sets \citep{schallhorn2017discrete} are daily and hourly records of bike rentals in Washington, D.C., together with local climate data (temperature, perceived temperature, humidity and wind speed). The hourly bike data set additionally contains the variable hour. The MAGIC (Major Atmospheric Gamma-ray Imaging Cherenkov) data has been analyzed in the context of classification by \cite{Nagler2015evading}. We only show the results for the subset with the classification hadron. Further, the Norwegian data is a financial data set of Norwegian and international market variables used in the context of truncated regular vines by \cite{brechmann2012truncated} and \cite{killiches2016using}. Two higher dimensional financial data sets are given by the CDS data \citep{brechmann2013conditional,kraus2017d} and the EuroStoxx 50 data \citep{brechmann2013risk,killiches2016using}. Finally, in order to include another non-financial data set, we also consider a subset of the Concrete Compressive Strength Data Set \citep{yeh1998modeling}, containing all of its continuous variables, namely \textit{Cement}, \textit{Coarse aggregate}, \textit{Fine aggregate} and \textit{Concrete compressive strength}.\\ 
The dimensions of the data sets range between 3 and 52 with sample sizes between 655 and 17379. All of the data was transformed to the copula scale using adequate preprocessing methods where necessary \citep[e.g.\ GARCH models for time series data as in][]{liu2009efficient} and empirical probability integral transforms. \autoref{tab:realdata} shows the AIC-values of the vine copulas fitted by the three competing algorithms. We further provide in brackets for each vine copula fitted by the three algorithms the number of pair-copulas for which the test for constant conditional correlations is rejected at the 5\% level.
So, e.g.\ in the case of the 3-dimensional uranium data set the CCC test is rejected with a p-value of 0.01 for the conditional copula modeled by Di\ss mann's algorithm (yielding a ``1'' in brackets) while \autoref{Alg2} chooses a structure with no rejection (implying the ``0'').

\begin{table}[ht]
	\begin{tabular}{c|ccccc}
		Data & uranium & concrete & bike daily & bike hourly & uranium\\
		\hline
		$d$ & 3 & 4 & 5 & 6 & 7\\
		\hline
		$n$ & 655 & 1030 & 731 & 17379 & 655\\
		\hline
		Di\ss mann & $-845.6$ (1) & $-522.9$ (2) & $-4423.5$ (3) & $-89859.6$ (9) & $-1759.9$ (4)\\ 
		Algorithm 1 & $-845.6$ (1) & $-522.9$ (2) & $-4423.5$ (3) & $-89842.4$ (9) & $-1756.4$ (4)\\ 
		Algorithm 2 & $\mathbf{-857.8}$ (0) & $\mathbf{-551.6}$ (3) & $\mathbf{-4454.1}$ (3) & $\mathbf{-92040.5}$ (10) & $\mathbf{-1817.2} (1)$\\\Xhline{4\arrayrulewidth}
		\multicolumn{5}{c}{}\\
	\end{tabular}
	\begin{tabular}{c|cccc}
		Data & MAGIC & Norwegian & CDS & EuroStoxx\\
		\hline
		$d$ & 10 & 19 & 38 & 52 \\
		\hline
		$n$ & 6688 & 1187 & 1371 & 985 \\
		\hline
		Di\ss mann & $-61359.4$ (23) & $-12906.4$ (18) & $\mathbf{-40985.7}$ (31) & $\mathbf{-62377.8}$ (87) \\ 
		Algorithm 1 & $\mathbf{-61649.7} (24)$ & $\mathbf{-12946.6}$ (10) & $-40922.4$ (25) & $-62280.1$ (65) \\ 
		Algorithm 2 & $-58197.4$ (17) & $-12756.5$ (12) & $-40514.5$ (32) & $-62074.8$ (68) \\\Xhline{4\arrayrulewidth}
	\end{tabular}
	\caption{AIC-values of the three structure selection methods applied to real data sets. In brackets the number of pair-copulas is given for which the CCC test is rejected at 5\% level.}
	\label{tab:realdata}
\end{table}

Similar to its three-dimensional subset the complete seven-dimensional uranium data set is best fitted by \autoref{Alg2}. Out of the 15 conditional pair-copulas of the vine fitted by Di\ss mann's algorithm four are deemed non-simplified at 5\% level. \autoref{Alg2} is able to reduce this number to one, thus yielding a better model fit. Regarding the four-dimensional concrete data set we see that \autoref{Alg2} is also able to improve the model fit. It may seem surprising, that compared to the other methods it increases the number of pair-copulas that violate the simplifying assumption. However, if we have a closer look at the p-values of the fitted pair-copulas, we see that the Di\ss mann fitted vine contains one pair-copula with p-value equal to zero which apparently severely impairs the model's likelihood. Similar observations can be made for the two bike data sets.\\
Considering the higher dimensional data sets we notice that \autoref{Alg2} looses its competitive advantage. The ten-dimensional MAGIC data as well as the 19-dimensional Norwegian data achieve their best fit using \autoref{Alg1}. For the MAGIC data set it increases the overall non-simplified pair-copulas while reducing the number of strong violations (with p-value equal to zero) from 12 to 8. The good performance of Di\ss mann's algorithm for the high-dimensional finance data sets is not surprising. It is a stylized fact that these are modeled fairly well by t copulas \citep[see for example][]{demarta2005t}. \cite{stoeber2013simplified} have shown that vine decompositions of t copulas are of the simplified form, independently of the vine's tree structure. Thus, the CCC test is rejected less often for vines modeling financial data. So for the 38-dimensional CDS data set, in the Di\ss mann vine only 31 of the 666 fitted pair-copulas violate the simplifying assumption, which is slightly less than the 5\% that we would expect if the null hypothesis would be true for all pair-copulas. Thus, for structure selection purposes the p-value score is rather unimportant in this example. Consequently, Di\ss mann's heuristic of modeling the strongest dependencies in the lower trees yields the best results and the restriction to C-vines imposed by \autoref{Alg2} is apparently too severe. Similar arguments hold for the 52-dimensional EuroStoxx data set, where only 6.8\% of the pair-copulas fitted by Di\ss mann's algorithm violate the simplifying assumption.\\
Regarding the pair-copula families selected by the three algorithms, we observe that they do not differ considerably between the algorithms. Not surprisingly, most of the pair-copula families of the financial data sets are t-copulas, with some Tawn, Frank and BB1 copulas in the mix. The concrete data was modeled mainly by Tawn, Frank and Gauss copulas. The same copula families were chosen for the two bike data sets, with some additional BB8 copulas and rotations thereof. For the uranium data set, Frank, Joe, t and Tawn copulas were favored, while the most commonly selected families of the MAGIC data were Tawn, BB8 and t copulas. So we see that in these examples many non-Gaussian copulas are selected such that Gaussian copulas would not provide satisfying model fits.\\
All in all, we have seen that the algorithms proposed in this paper are able to improve the model fit compared to Di\ss mann's algorithm, especially when the Di\ss mann vine exhibits significant non-simplifiedness as assessed by the CCC test.

\section{Conclusion}\label{sec:conclusion}

We propose two new algorithms for the sequential selection of the tree structure of a vine copula model. We extend currently existing methods by incorporating tests which gage the validity of the simplifying assumption for every pair-copula. This results in vine copula models which frequently have a better model fit than the benchmark given by Di\ss mann's algorithm.\\
We have revisited the three-dimensional uranium data set which is famous for its non-simplified nature when considering its vine decomposition using Di\ss mann's algorithm. However, we found out that using the vine decomposition where the conditioning in the second tree is done with respect to cobalt, the resulting vine is of the simplified form and provides a better model fit.\\
In the real data application we have seen that our proposed algorithms work especially well, when the vine fitted by Di\ss mann's algorithm contains many pair-copulas violating the simplifying assumption. Thus, from a practitioner's point of view, after fitting a $d$-dimensional vine copula using Di\ss mann's algorithm one should always count the number of pair-copulas for which the test of constant conditional correlations is rejected at some level $\beta$. If this number is considerably larger than $\beta(d-1)(d-2)/2$  (the expected number of rejections), one should refit the model using our proposed algorithms, accepting larger computational times as a trade-off for a likely better model fit.\\

\section*{Acknowledgments}\label{sec:acknowledgment}
This work was supported by the German Research Foundation (DFG grant CZ 86/4-1). The authors would like to thank Malte Kurz, Fabian Spanhel and Thomas Nagler for useful comments and fruitful discussions.

\section*{Appendix}
\begin{appendix}
\section{Pair-copula families implemented in \texttt{VineCopula}}\label{app:families}	
The pair-copula families currently implemented in the \texttt{VineCopula} package are Gaussian, t*, Clayton, Gumbel, Frank, Joe, BB1*, BB6*, BB7*, BB8* and Tawn* with respective rotations. The stars indicate two-parametric families.

\section{Simulation of the second parameter for two-parametric copula families}\label{app:secondpar}
\paragraph*{t copula} The degrees of freedom of the t copula are sampled from $3+G$, where $G\sim Gamma(3,3)$. This implies an expected value of 4 and the 95\% confidence interval $[3.2,5.4]$.
\paragraph*{BB1, BB6, BB7} The second parameters of the BB1, BB6 and BB7 copulas are sampled from $1+3B$, where $B\sim Beta(4,2)$. This implies an expected value of 3 and the 95\% confidence interval $[1.84,3.85]$. When the dependence is negative, the second parameter is multiplied with $-1$.
\paragraph*{BB8, Tawn} The second parameters of the BB8 and Tawn copulas are sampled from a $Beta(4,2)$ distribution. This implies an expected value of $2/3$ and the 95\% confidence interval $[0.28,0.95]$. When the dependence is negative, the second parameter of the BB8 copula is multiplied with $-1$.
\section{Choice of weighting factor $\alpha$}\label{app:choiceofalpha}
\begin{figure}[!htb]
	\centering
	\includegraphics[trim=0cm 0cm 0cm 0cm,clip,width=0.32\textwidth]{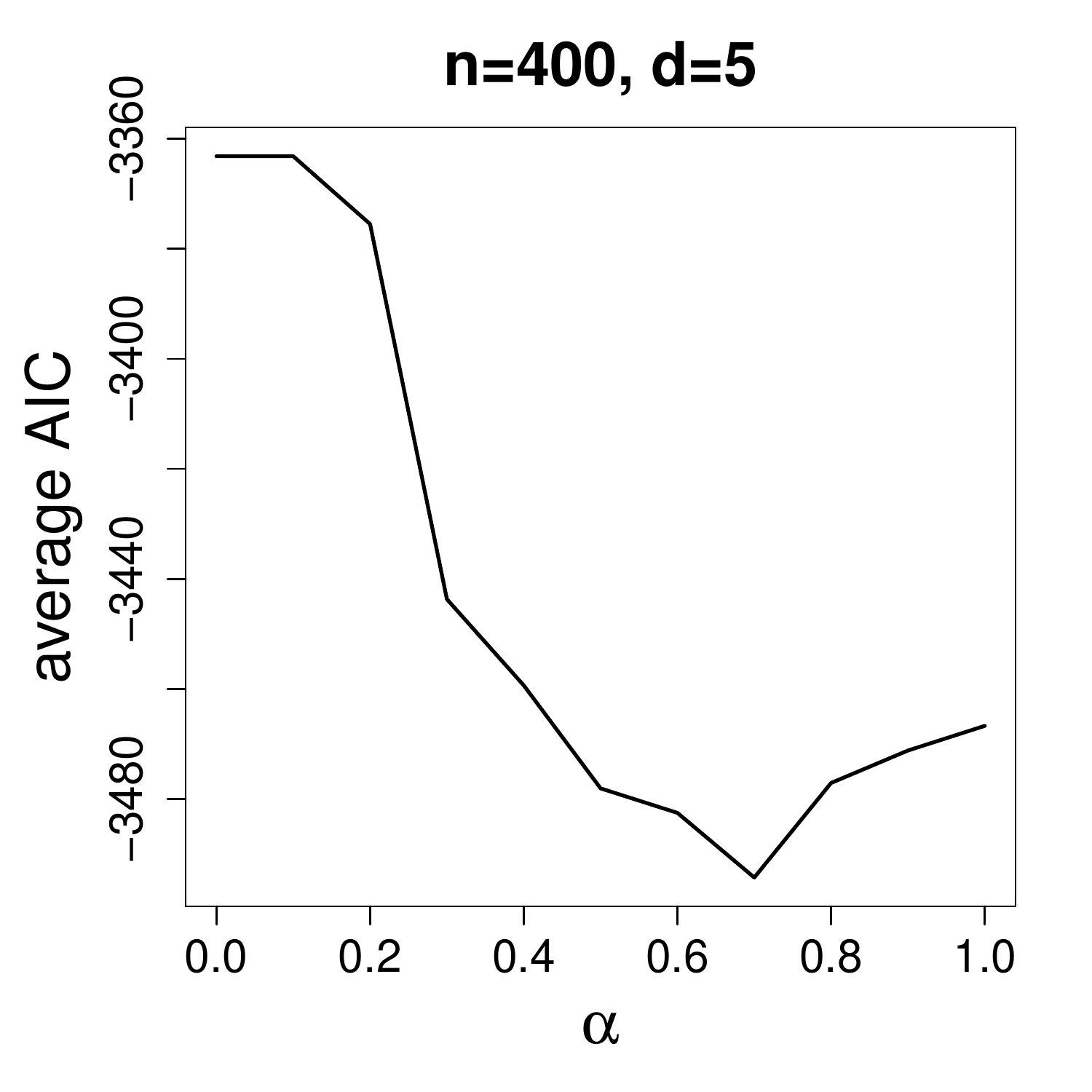}
	\includegraphics[trim=0cm 0cm 0cm 0cm,clip,width=0.32\textwidth]{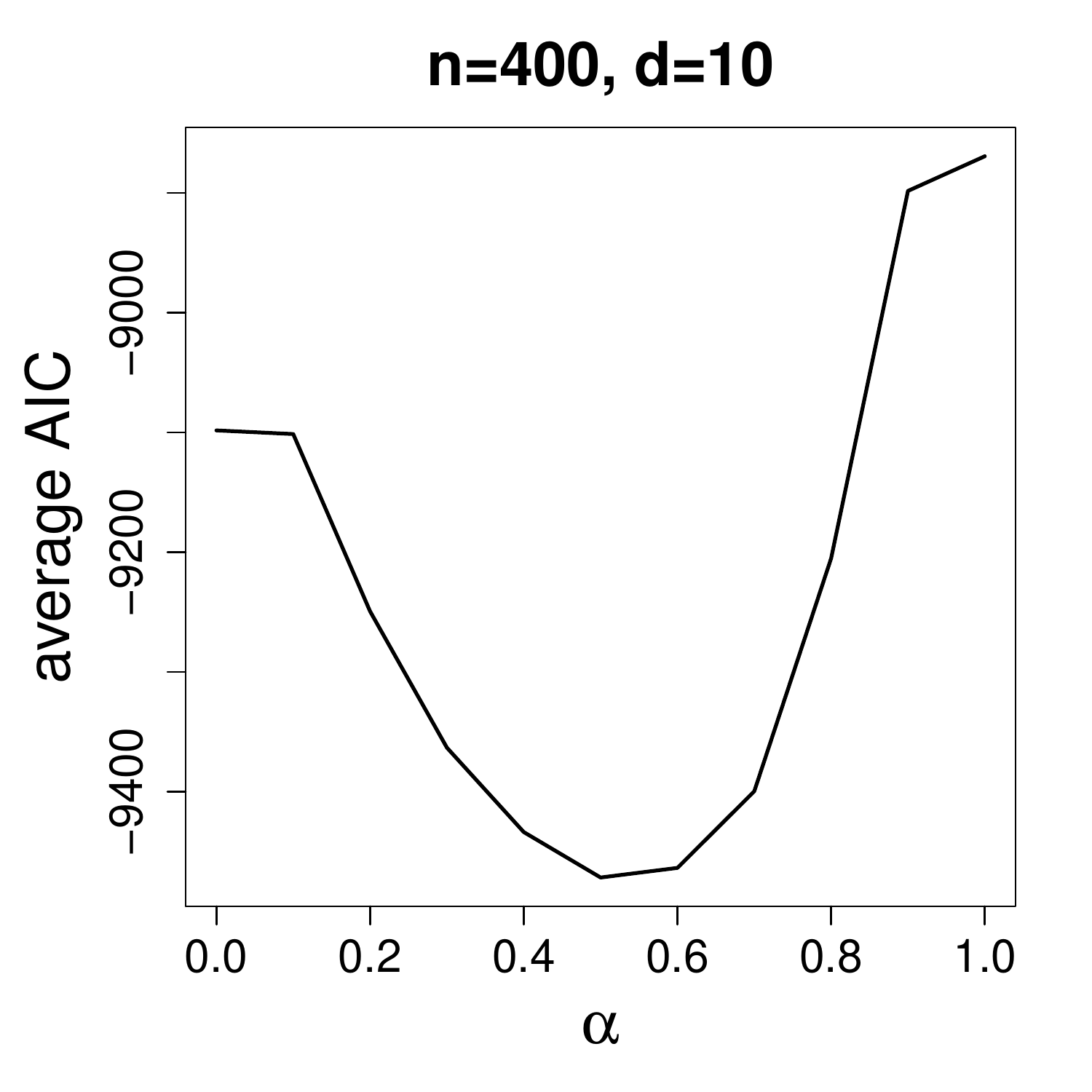}
	\includegraphics[trim=0cm 0cm 0cm 0cm,clip,width=0.32\textwidth]{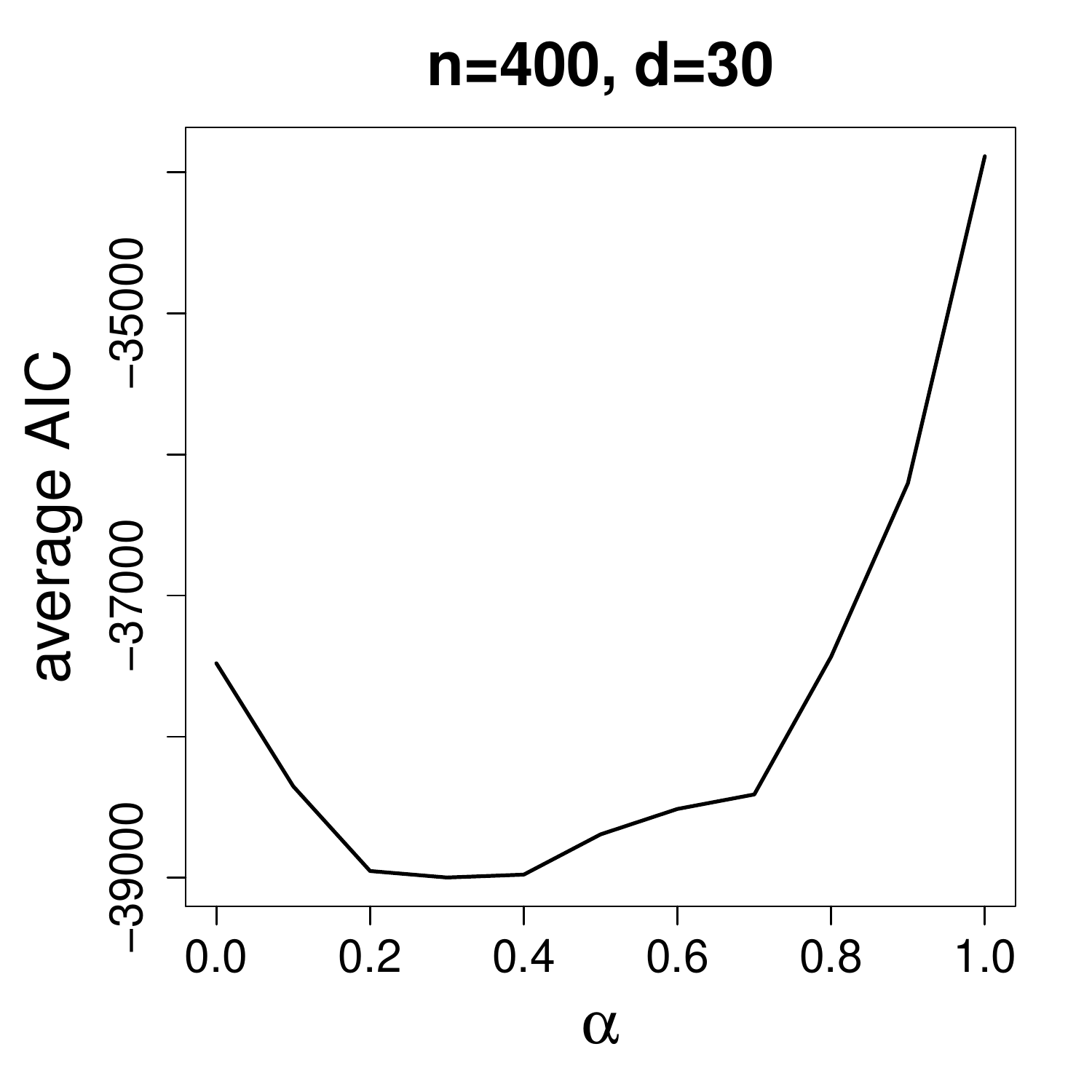}
	\includegraphics[trim=0cm 0cm 0cm 0cm,clip,width=0.32\textwidth]{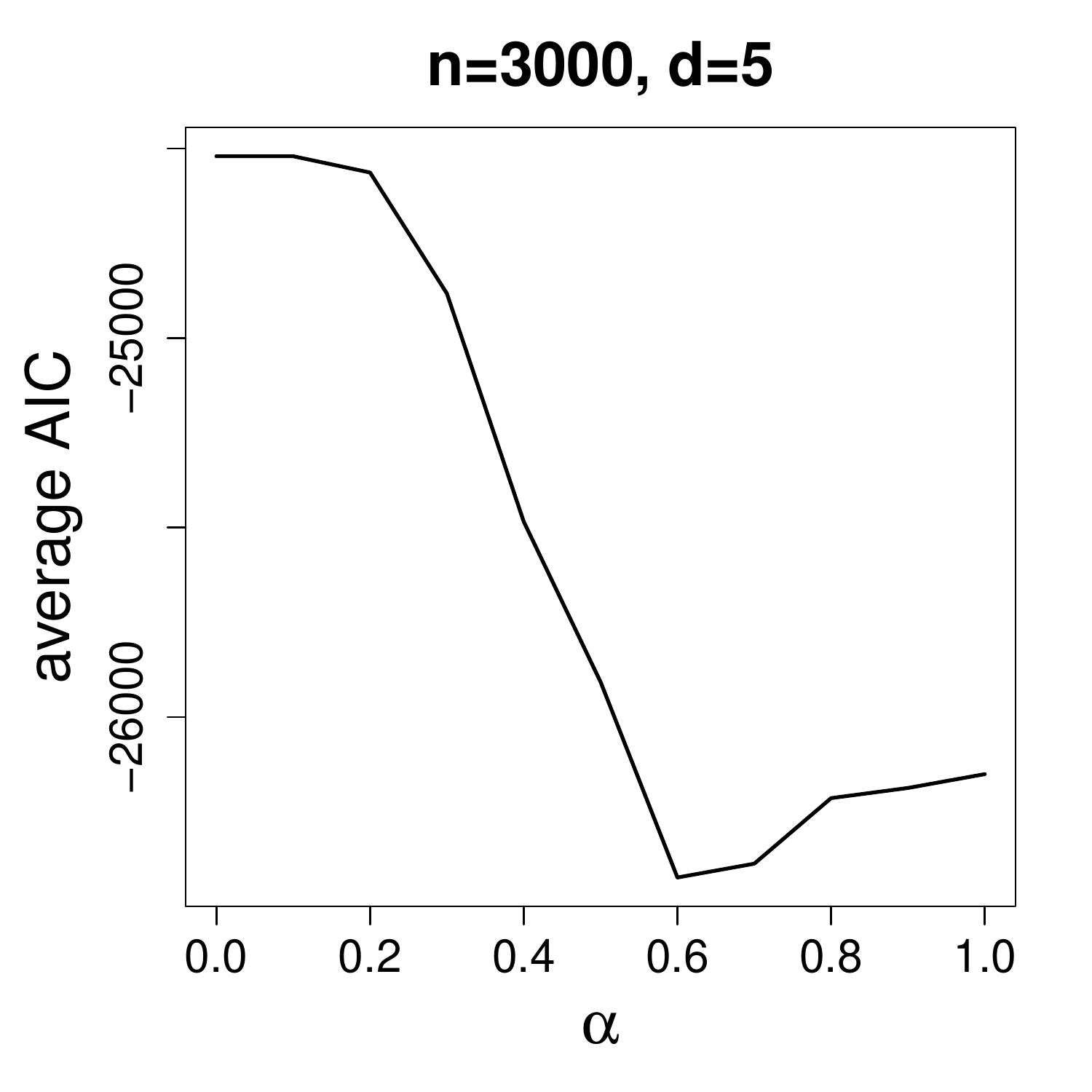}
	\includegraphics[trim=0cm 0cm 0cm 0cm,clip,width=0.32\textwidth]{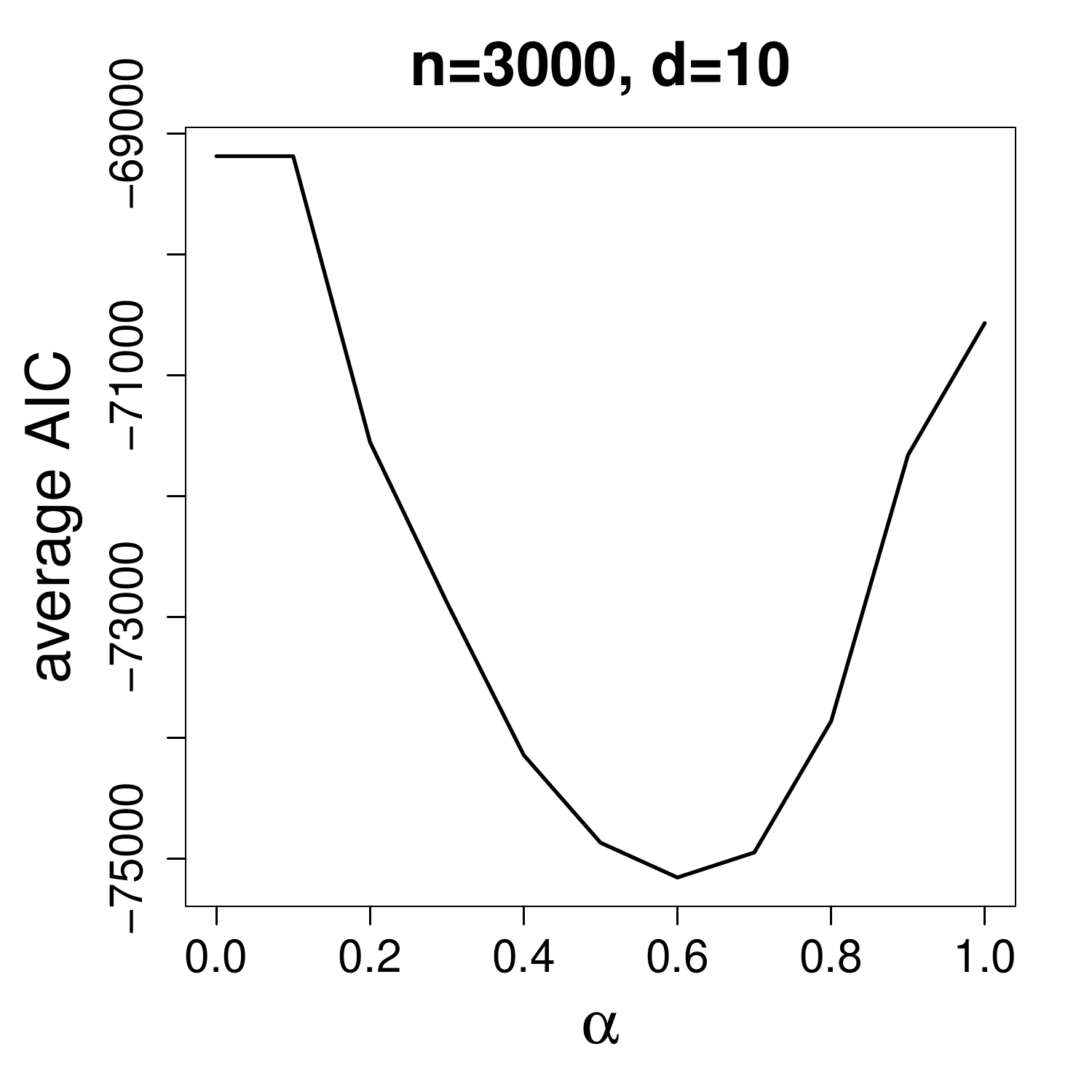}
	\includegraphics[trim=0cm 0cm 0cm 0cm,clip,width=0.32\textwidth]{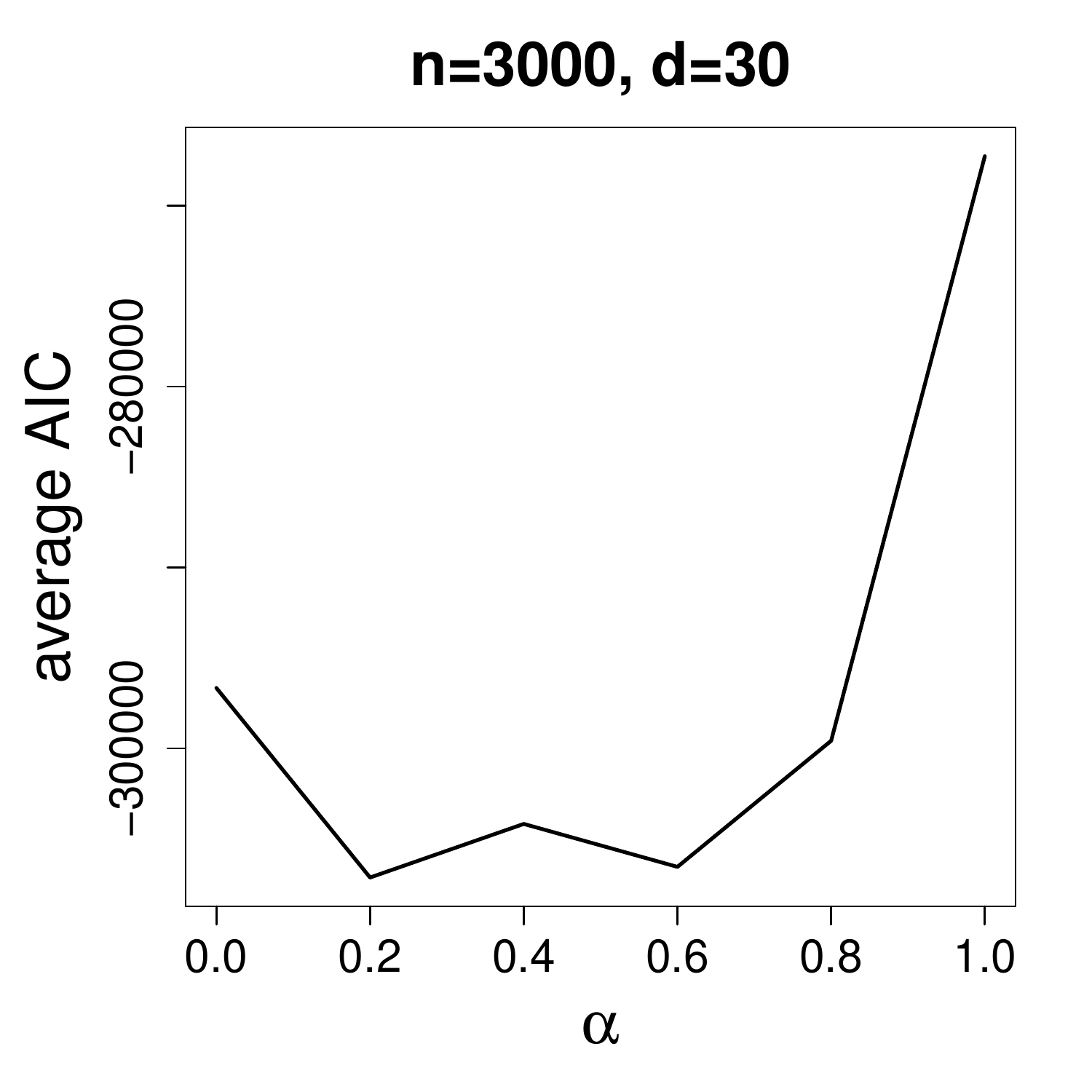}
	\caption{AIC-values of vine copulas fitted by \autoref{Alg2} depending on the weighting factor $\alpha$ in the settings $n=400$ (top row) and $n=3000$ (bottom row) for $d=5,10,30$, averaged over 1000 repetitions.}
	\label{fig:choiceofalphaApp}
\end{figure}

\newpage
\section{Differences of AIC-values}\label{app:boxplots}
\begin{figure}[!htb]
	\centering
	\includegraphics[trim=0cm 1.5cm 0cm 1.7cm,clip,width=0.48\textwidth]{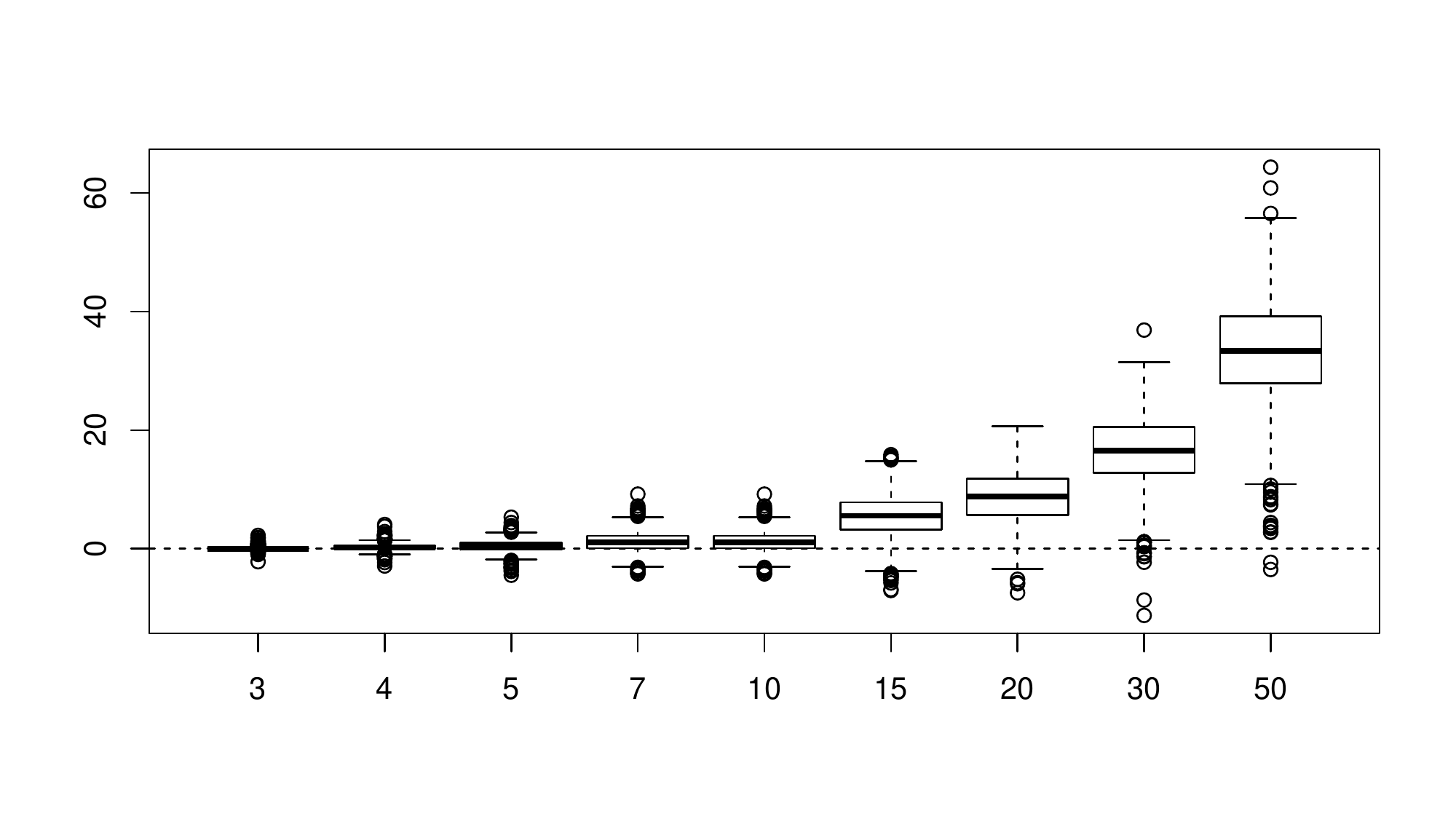}
	\includegraphics[trim=0cm 1.5cm 0cm 1.7cm,clip,width=0.48\textwidth]{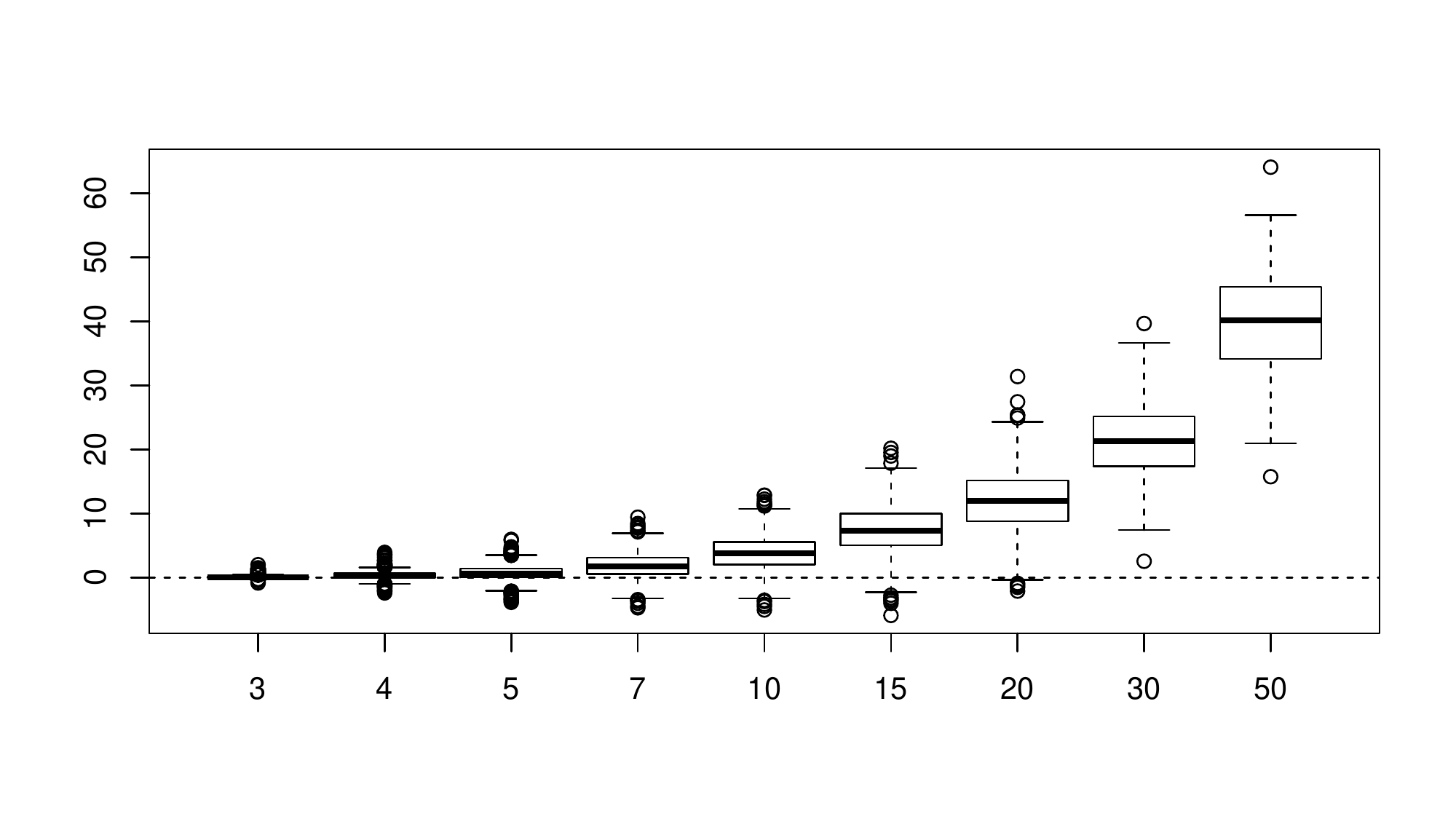}
	\caption{Boxplots of the difference between the AIC-values per observation of the vines chosen by Di\ss mann's algorithm and \autoref{Alg2} for $n=400$ (left panel) and $n=3000$ (right panel).}
	\label{fig:AIC_boxplot2}
\end{figure}

	\section{Simulation study results for positive dependence and independence tests}\label{app:posdep_ind}
	
	\begin{table}[ht]
		\centering
		\begin{tabular}{c|c|llll}
			&&\multicolumn{4}{c}{$d$}\\
			\hline
			Ind. test & Algorithm & 3 & 5 & 10  & 30 \\ 
			\hline
			\multirow{2}{*}{Without} & Algorithm 1 & 100 (100) & 73.1 (53.4) & 37.0 (3.2) & 18.8\\ 
			& Algorithm 2 & 94.1 (22.9) & 83.1 & 86.1 & 91.9 \\			
			\hline
			\multirow{2}{*}{With} & Algorithm 1 & 100 (100) & 73.6 (54.0) & 33.0 (3.2) & 16.8 \\
			& Algorithm 2 & 94.4 (22.9) & 82.9 & 85.8 & 92.9 \\\Xhline{4\arrayrulewidth}
		\end{tabular}
		\caption{Percentages of better or equal performance regarding the AIC-value of the two algorithms compared to Di\ss mann's algorithm for $n=1000$ and random vine copulas with only positive dependence (in brackets the percentages of equal performance are given). In the first two rows pair-copula no independence tests were performed and in the last two rows they were performed with level $\beta=0.05$.}
		\label{tab:simstudy_pos_ind}
	\end{table}
	
\section{Choice of p-value transformation function $r$}\label{sec:choice_r}
\begin{table}[ht]
	\centering
	\begin{tabular}{c|llll}
		&\multicolumn{4}{c}{$d$}\\
		\hline
		$r$ & 3 & 5 & 10  & 30 \\ 
		\hline
		rank      & 91.1 (51.5) & 73.6 & 88.7 & 99.6 \\
		identity  & 89.9 (48.5) & 71.1 & 87.5 & 99.5 \\
		logarithm & 89.9 (48.5) & 72.4 & 77.0 & 94.8\\\Xhline{4\arrayrulewidth}
	\end{tabular}
	\caption{Percentages of better or equal performance regarding the AIC-value of \autoref{Alg2} compared to Di\ss mann's algorithm depending on the transformation function $r$ for $n=1000$ and $d=3,5,10,30$ (in brackets the percentages of equal performance are given).}
	\label{tab:choice_r}
\end{table}
\newpage
\section{Plots of simulation study results}\label{sec:plot_simstudy}

\begin{figure}[!htb]
	\centering
	\includegraphics[trim=0cm 0cm 0cm 0cm,clip,width=0.32\textwidth]{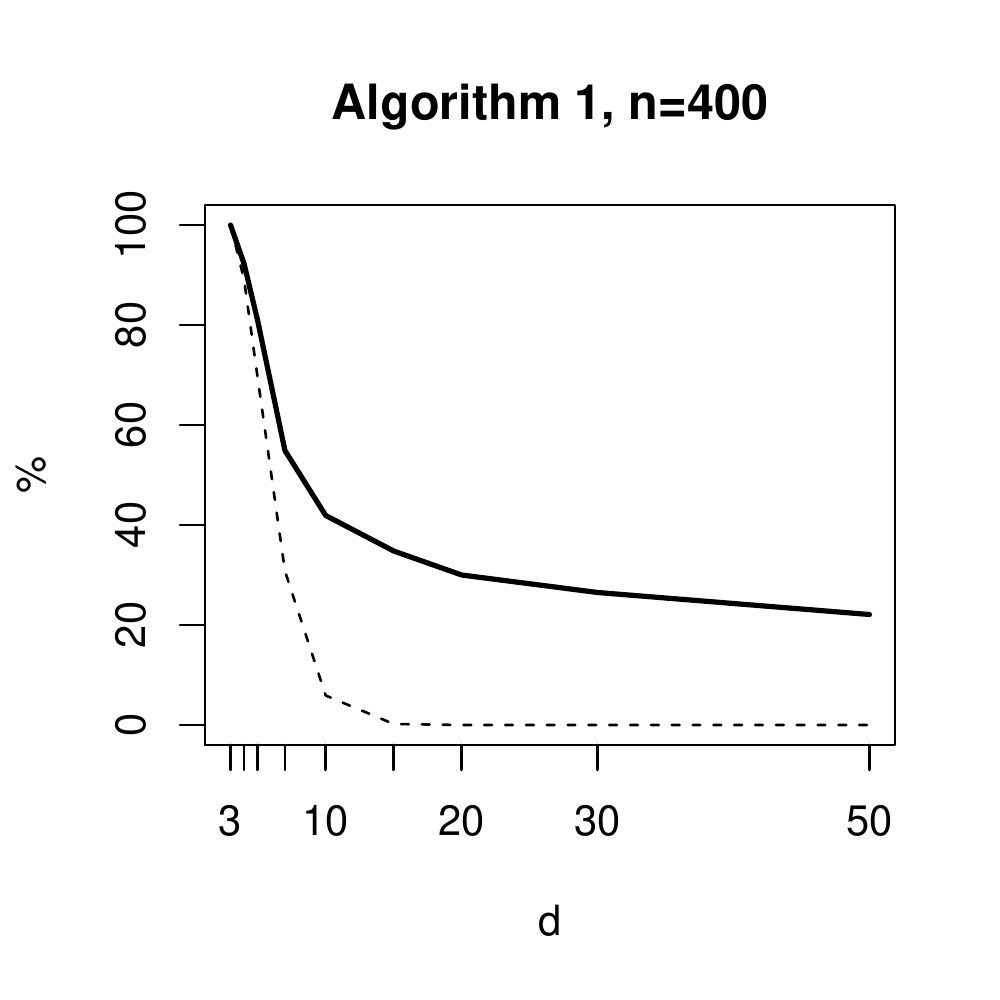}
	\includegraphics[trim=0cm 0cm 0cm 0cm,clip,width=0.32\textwidth]{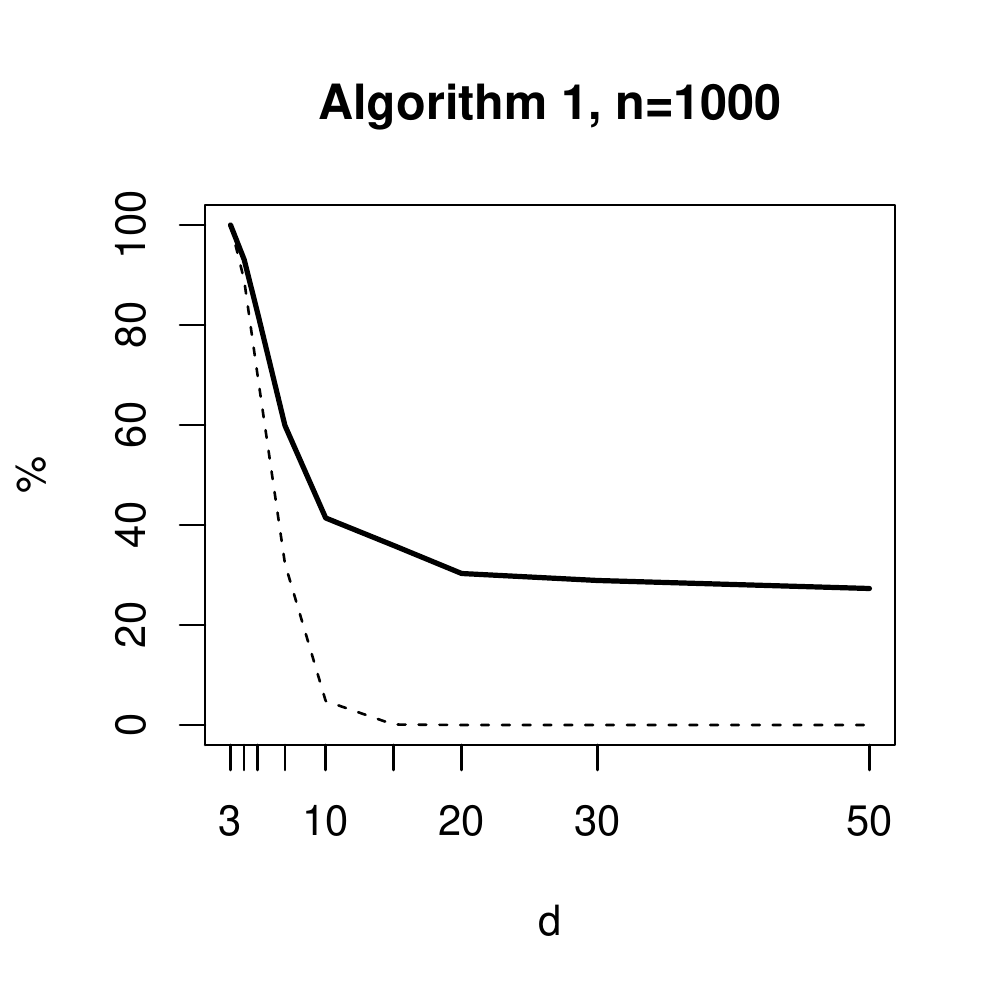}
	\includegraphics[trim=0cm 0cm 0cm 0cm,clip,width=0.32\textwidth]{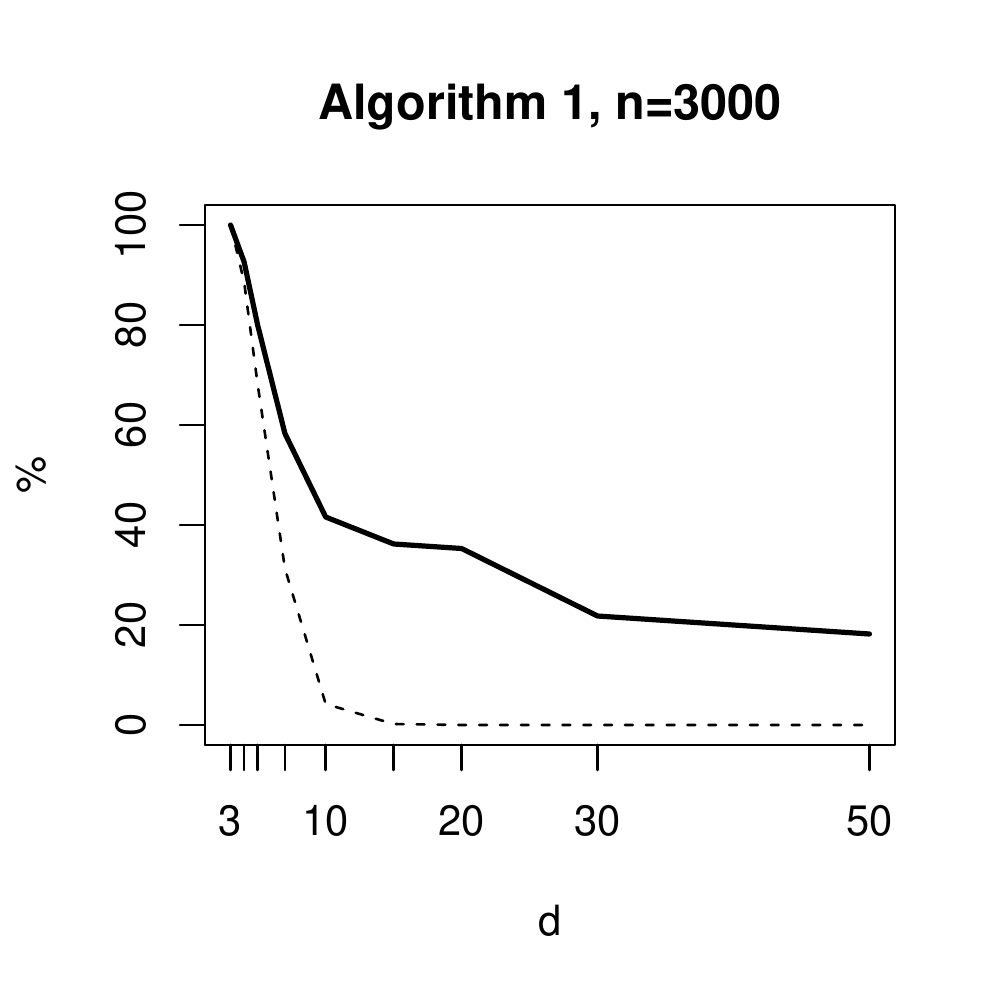}
	\includegraphics[trim=0cm 0cm 0cm 0cm,clip,width=0.32\textwidth]{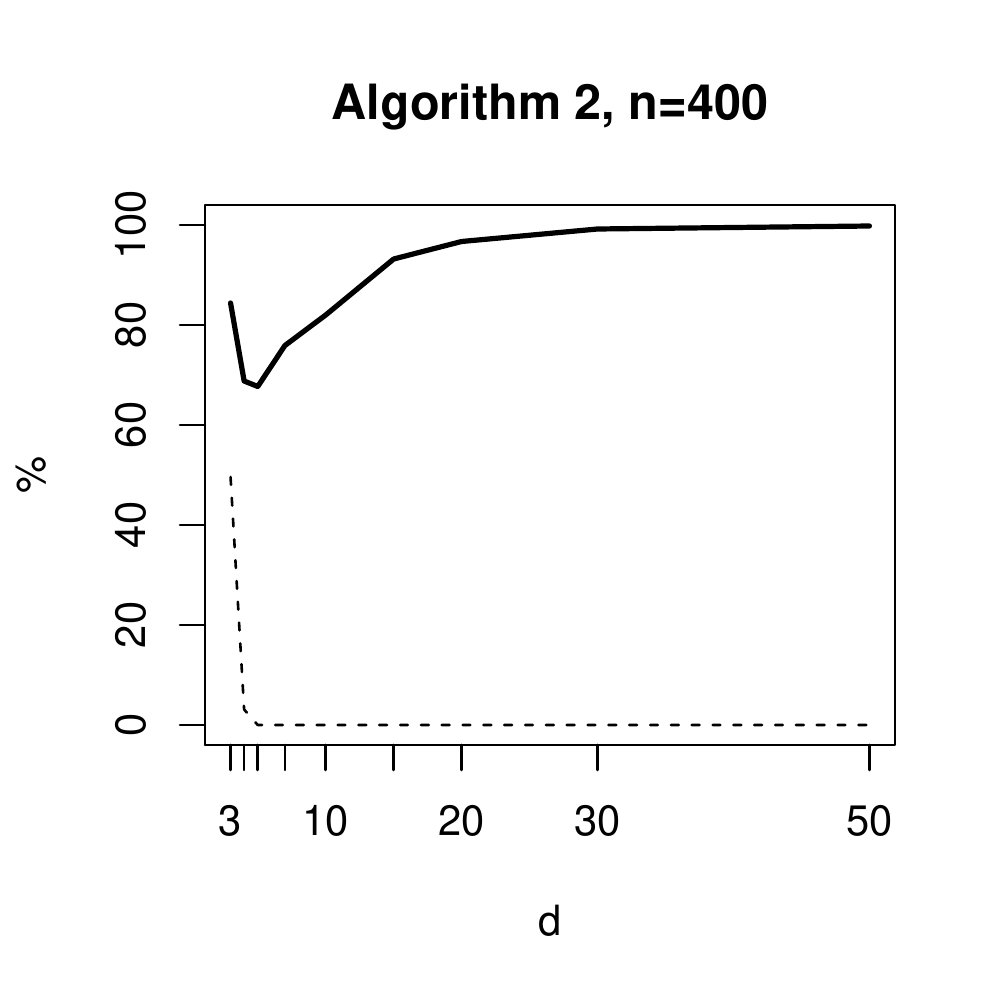}
	\includegraphics[trim=0cm 0cm 0cm 0cm,clip,width=0.32\textwidth]{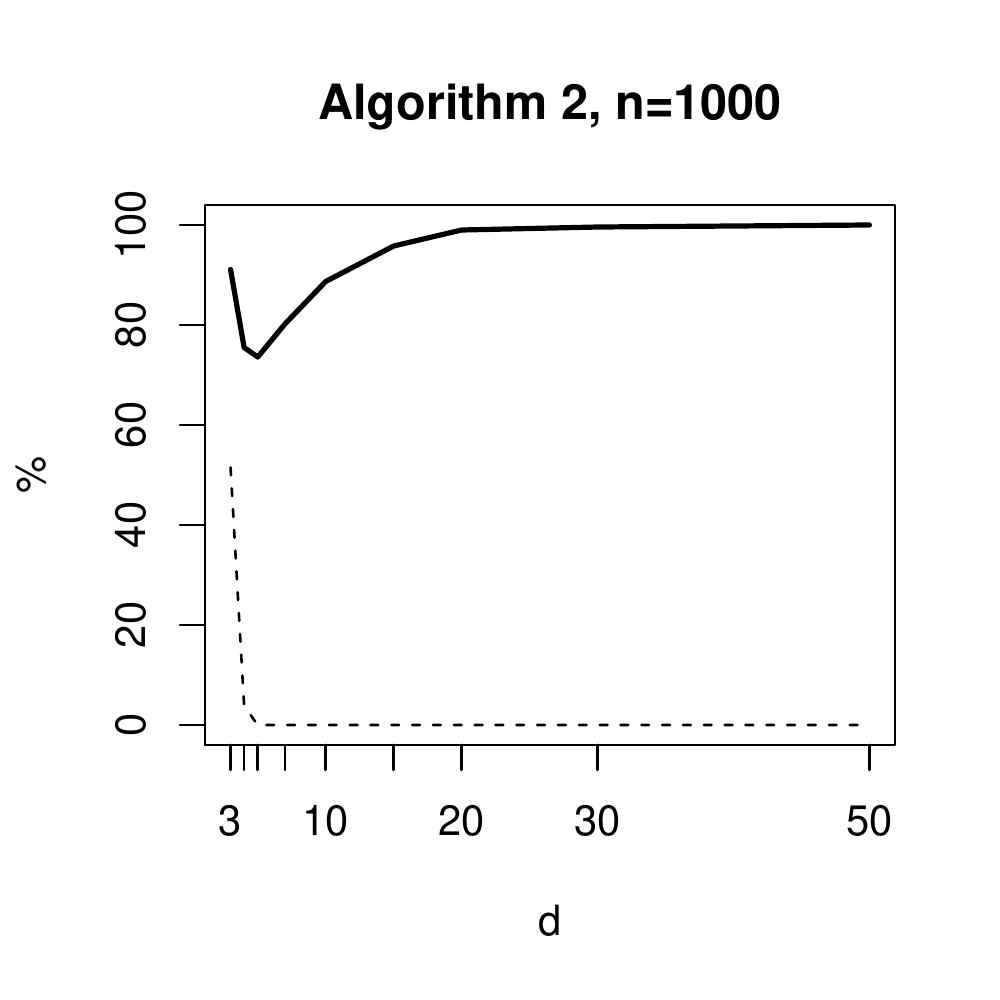}
	\includegraphics[trim=0cm 0cm 0cm 0cm,clip,width=0.32\textwidth]{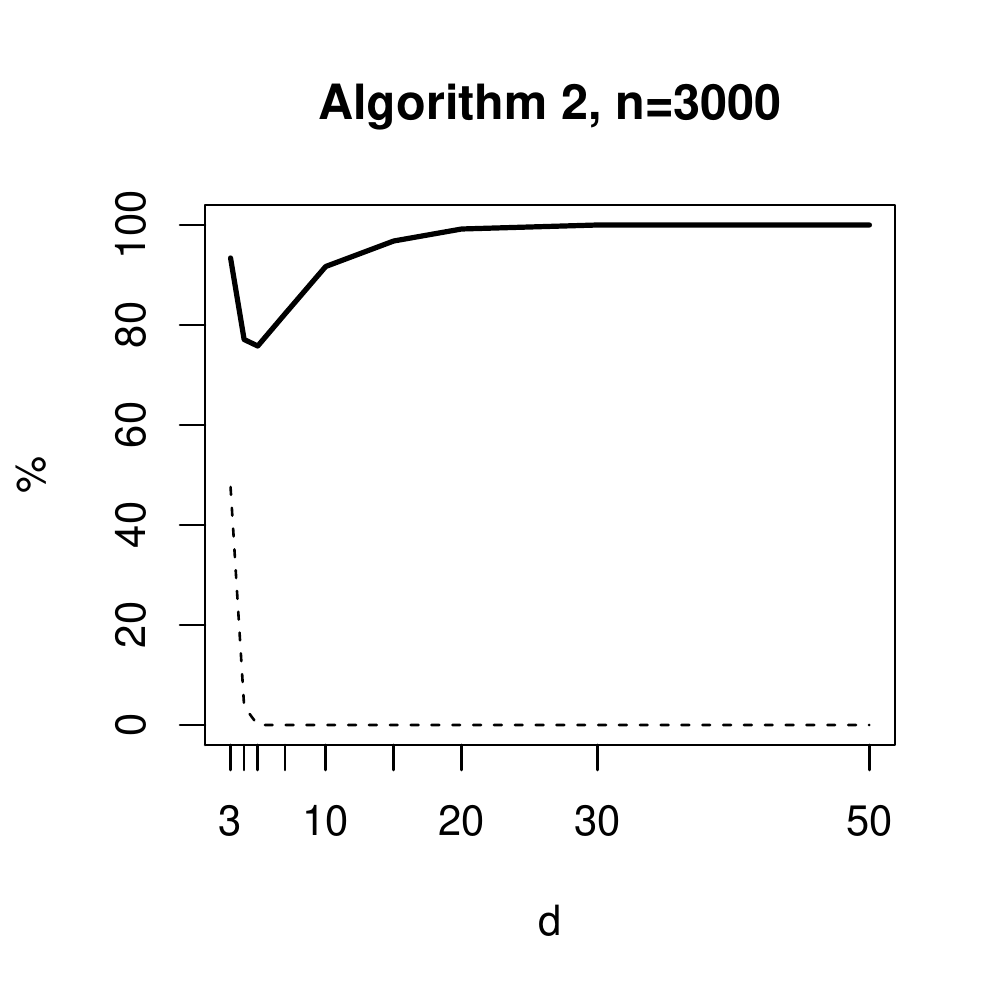}
	\caption{Percentages of better or equal performance regarding the AIC-value of the two algorithms compared to Di\ss mann's algorithm for different dimensions $d$ and sample sizes $n$ based on 1000 data sets sampled from randomly generated R-vines (the dashed lines represent the percentages of equal performance).}
	\label{fig:simstudy_results}
\end{figure}

\end{appendix}

\bibliography{References}{}
\bibliographystyle{asa}

\end{document}